\documentclass[aps,prab,twocolumn,superscriptaddress]{revtex4-1}

\usepackage{graphicx}
\usepackage{enumitem}

\usepackage{gensymb}

\begin{document}

\title{A fast-switching magnet serving a spallation-driven ultracold neutron source}

\newcommand{\UofM}{\affiliation{University of Manitoba, Winnipeg, MB R3T 2N2, Canada}}
\newcommand{\UBC}{\affiliation{The University of British Columbia, Vancouver, BC V6T 1Z1, Canada}}
\newcommand{\CERN}{\affiliation{Center for European Nuclear Research, CH-1211 Geneva 23, Switzerland}}
\newcommand{\McGill}{\affiliation{McGill University, Montreal, QC H3A 0G4, Canada}}
\newcommand{\TRIUMF}{\affiliation{TRIUMF, Vancouver, BC V6T 2A3, Canada}}
\newcommand{\UofW}{\affiliation{The University of Winnipeg, Winnipeg, Manitoba R3B 2E9, Canada}}
\newcommand{\Coburg}{\affiliation{Coburg University of Applied Sciences and Arts, 96450 Coburg, Germany}}
\newcommand{\RCNP}{\affiliation{Research Center for Nuclear Physics, Osaka University, Osaka 567-0047, Japan}}
\newcommand{\KEK}{\affiliation{High Energy Accelerator Research Organization, KEK, Tsukuba 305-0801, Japan}}
\newcommand{\Kyoto}{\affiliation{Kyoto University, Kyoto 606-8501, Japan}}
\newcommand{\UNBC}{\affiliation{University of Northern British Columbia, Prince George, BC V2N 4Z9, Canada}}
\newcommand{\SFU}{\affiliation{Simon Fraser University, Burnaby, BC V5A 1S6, Canada}}
\newcommand{\RIKEN}{\affiliation{RIKEN, Wako, Saitama 351-0198, Japan}}

\author{S.~Ahmed}
\UofM
\author{E.~Altiere}
\UBC
\author{T.~Andalib}
\UofM
\author{M.J.~Barnes}
\CERN
\author{B.~Bell}
\TRIUMF
\McGill
\author{C.P.~Bidinosti}
\UofW
\UofM
\author{Y.~Bylinsky}
\TRIUMF
\author{J.~Chak}
\TRIUMF
\author{M.~Das}
\UofM
\author{C.A.~Davis}
\TRIUMF
\author{F.~Fischer}
\TRIUMF
\Coburg
\author{B.~Franke} 
\TRIUMF
\author{M.T.W.~Gericke} 
\UofM
\author{P.~Giampa} 
\TRIUMF
\author{M.~Hahn}  
\TRIUMF
\Coburg
\author{S.~Hansen-Romu} 
\UofM
\author{K.~Hatanaka} 
\RCNP
\author{T.~Hayamizu} 
\UBC
\RIKEN
\author{B.~Jamieson} 
\UofW
\UofM
\author{D.~Jones} 
\UBC
\author{K.~Katsika} 
\TRIUMF
\author{S.~Kawasaki} 
\KEK
\author{T.~Kikawa} 
\TRIUMF
\Kyoto
\author{W.~Klassen} 
\UofM
\author{A.~Konaka} 
\TRIUMF
\author{E.~Korkmaz} 
\UNBC
\author{F.~Kuchler}  
\TRIUMF
\author{L.~Kurchaninov} 
\TRIUMF
\author{M.~Lang} 
\UofM
\author{L.~Lee} 
\TRIUMF
\author{T.~Lindner} 
\TRIUMF
\UofW
\author{K.W.~Madison} 
\UBC
\author{J.~Mammei} 
\UofM
\author{R.~Mammei} 
\UofW
\TRIUMF
\UofM
\author{J.W.~Martin} 
\UofW
\UofM
\author{R.~Matsumiya} 
\TRIUMF
\author{E.~Miller} 
\UBC
\author{T.~Momose} 
\UBC
\author{R.~Picker}
\TRIUMF
\SFU
\author{E.~Pierre} 
\TRIUMF
\RCNP
\author{W.D.~Ramsay} 
\TRIUMF
\author{Y.-N.~Rao} 
\TRIUMF
\author{W.R.~Rawnsley} 
\TRIUMF
\author{L.~Rebenitsch} 
\UofM
\author{W.~Schreyer} 
\TRIUMF
\author{S.~Sidhu} 
\SFU
\TRIUMF
\author{S.~Vanbergen} 
\TRIUMF
\UBC
\author{W.T.H.~van~Oers} 
\TRIUMF
\UofM
\author{Y.X.~Watanabe} 
\KEK
\author{D.~Yosifov} 
\TRIUMF

\date{\today}

\begin{abstract}
  A fast-switching, high-repetition-rate magnet and power supply have been developed for and operated at
  TRIUMF, to deliver a proton beam to the new ultracold neutron (UCN) facility.
  The facility possesses unique operational requirements: a time-averaged beam current of 40~$\mu$A with the ability to switch the
  beam on or off for several minutes.
  These requirements are in conflict with the typical operation mode of the TRIUMF
  cyclotron which delivers nearly continuous beam to multiple users.
  To enable the creation of the UCN facility, a beam-sharing arrangement with
  another facility was made.  The beam sharing is accomplished by the
  fast-switching (kicker) magnet which is ramped
  in 50~$\mu$s to a current of 193~A, held there for approximately 1~ms, then ramped down in
  the same short period of time.  This achieves a 12~mrad deflection which is sufficient to switch the proton beam between the two facilities.  The kicker magnet relies on a
  high-current, low-inductance coil connected
  to a fast-switching power supply that is based
  on insulated-gate bipolar transistors (IGBTs).  The design and
  performance of the kicker magnet system and initial beam delivery results are reported.
\end{abstract}

\pacs{}

\maketitle

\section{Background motivation}

The TRIUMF cyclotron delivers a nearly continuous proton beam with a pulse frequency of 1.126~kHz.
The cyclotron accelerates H$^-$ ions, 
using stripper foils to deliver protons at various radii
(energies) to up to four beamlines (numbered 1, 2A, 2C, and 4) simultaneously.  The cyclotron
typically delivers a total current in excess of 300\,$\mu$A, at a
maximum energy of 520\,MeV.

A new ultracold neutron (UCN) source was
developed at TRIUMF, with the initial planning beginning in 2008.  As shall be described, the UCN source requires a minute-long beam-pulse structure.  
Accommodating an additional facility which depends on intermittent beam delivery at variable intensity required the ability to share the beam.  This was implemented in beamline 1 at TRIUMF and a beam-sharing arrangement made with the meson-production facilities.
Coexistence and simultaneous operation of the new UCN facility is only achievable due to the fast-switching kicker magnet which switches between the facilities.

Ultracold neutrons are slowly moving (less than 8\,m/s) neutrons
which may be contained in material, magnetic, or gravitational
traps for long periods of time (hundreds of seconds).  The new UCN source is
based on neutrons produced from a tungsten target via
proton-induced spallation, followed by moderation and superthermal UCN production in superfluid helium,
cooled to temperatures around 1\,K~\cite{GOLUB1975133,PhysRevC.92.024004,KOROBKINA2002462}.  Once produced, UCN may remain in the source
volume for long periods of time.  After switching the proton beam onto the spallation target, the UCN density asymptotically approaches its maximum value on a timescale of typically 60\,s, at which
point the proton beam may be switched off again and the UCN transported to
sensitive counting experiments.  This suggests that a proton beam that can be switched on and off for minutes at a time is desirable to build up UCN density within
the source and reduce background during the experiment counting period.

The larger the beam current on the spallation target, the more spallation neutrons are produced, but also the more heat is deposited in the superfluid helium by highly energetic spallation particles. However, the colder the superfluid helium, the more UCN can be accumulated in the source. Thus, an adjustable beam current is desirable to find the optimal operating point for a UCN source.  The facility was designed around the requirement of 483\,MeV protons at a maximum 40\,$\mu$A current being delivered to the spallation target.

The subject of this paper is the design, fabrication,
testing, commissioning, and successful operation of a kicker magnet system that uses a pulsed beam structure
to intermittently divert beam from beamline 1 at TRIUMF into a newly installed beamline 1U with the spallation target.
Several new diagnostic tools had to be created to ensure successful
operation and these are also described.  The system has been
operating reliably at TRIUMF since early 2017.

\section{Beam structure}

\begin{figure}
  \includegraphics[width=\linewidth]{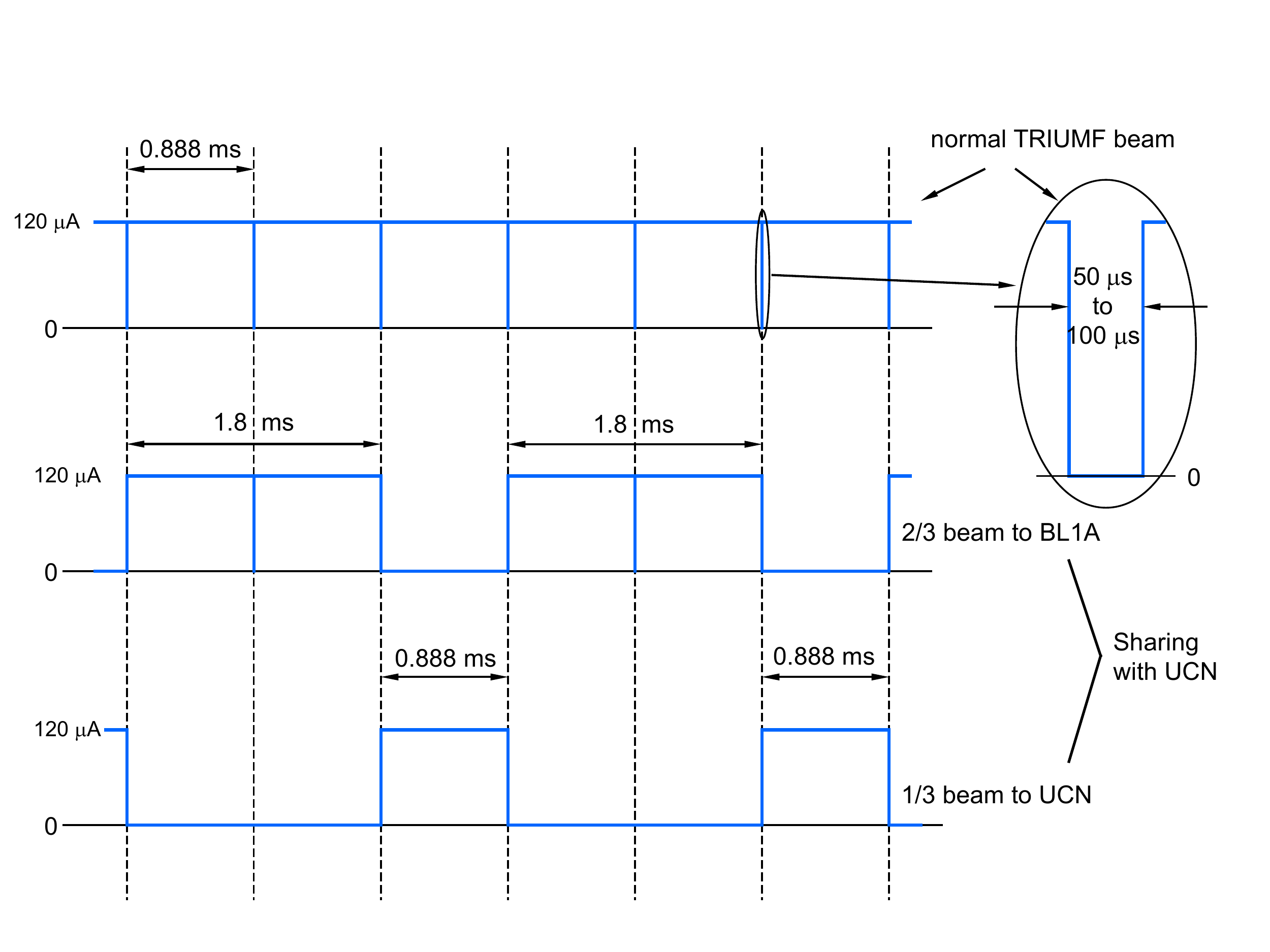}
  \caption{Sharing of the main beam from the cyclotron (normal TRIUMF
    beam) between beamline 1A and beamline 1U.
    Beam is diverted into beamline 1U by switching on the kicker magnet
    during the 50--100~$\mu$s notch duration, then switching it off
    during the next notch. The pulser period is 0.888~ms.}
  \label{fig:beam-sharing}
\end{figure}

The structure of the beam in the TRIUMF cyclotron is shown
schematically in Fig.~\ref{fig:beam-sharing}.  Every 0.888~ms, a gap, or
notch, of 50--100~$\mu$s exists.  The notch is generated prior to beam
injection to the cyclotron and is present for all simultaneously
accelerated beams.  The duration of the notch has been used as a
tuning parameter to make fine adjustments to the power deposition of the beam in
targets in the 2A beamline, where stable and controllable power is
necessary to generate high-intensity ion beams for the Isotope
Separator and Accelerator (ISAC) radioactive beams facility~\cite{ISACandARIEL}.

The UCN kicker system is installed in beamline 1 (Fig.~\ref{fig:1VUAschematic}). The proton beam is normally transported down beamline 1A through
two meson-production targets to a beam dump.  
The beam energy is
483\,MeV and an average current of 100--120\,$\mu$A is typically delivered (Fig.~\ref{fig:beam-sharing}).

When beam is desired in beamline 1U to produce spallation neutrons, the current in the
kicker magnet is ramped up within the notch, held at a
constant current of 193\,A, and then ramped down during the next notch (Fig.~\ref{fig:beam-sharing}). The period between notches is referred to as a beam pulse. The beam may be switched on and off in this fashion so that up to one in three beam pulses
are delivered to beamline 1U, resulting in a repetition rate of 375\,Hz and an average 1U beam current of 33--40\,$\mu$A. 
The remaining pulses are delivered to beamline 1A as usual. The time when the beam is being switched between 1A and 1U in this manner is referred to as a beam-on period or a kicking period.  
The beam-on period is typically 60\,s long and the beam-off period is typically 180~s, so that the meson-production targets receive an integrated current 90\% of the original value.
The system
can also be operated in continuous (dc) mode, diverting all pulses into beamline 1U, or at lower repetition rates, resulting in a lower average current. Typical field rise/fall times for kicker systems range from tens to hundreds of nanoseconds and pulse widths generally 
range from tens of nanoseconds to tens of microseconds~\cite{Barnes:CAS2017}. The ability to operate in dc mode is an unusual requirement for a kicker system.

\section{Beamline layout}

\begin{figure*}
    \includegraphics[angle=90,width=\textwidth]{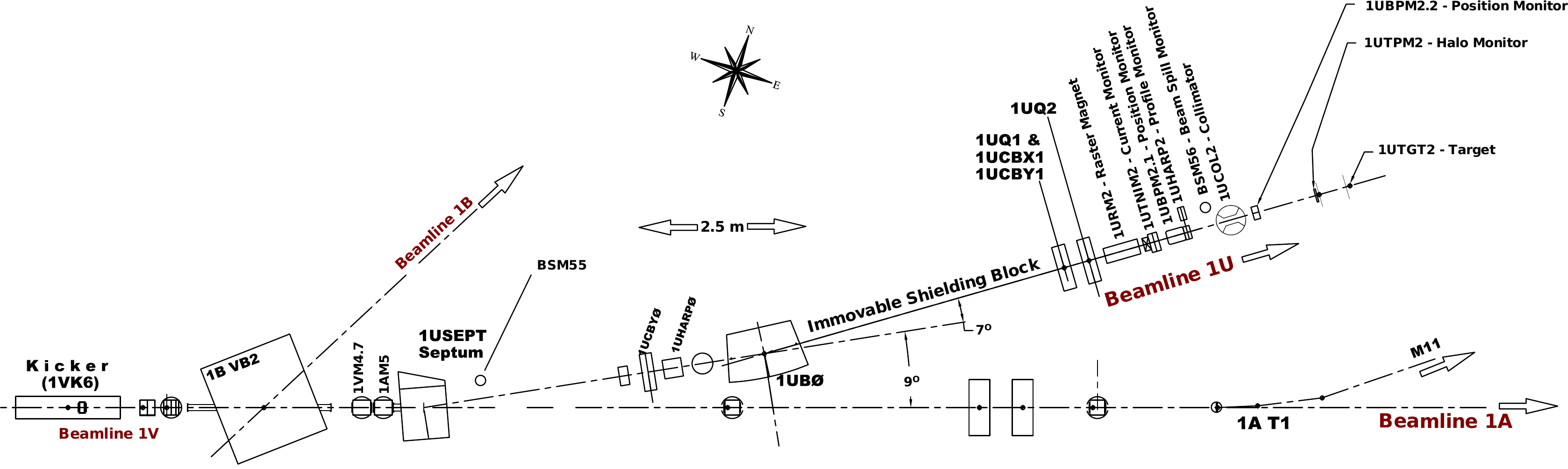}
\caption{A schematic top view of the proton beamline for the UCN facility at TRIUMF (see also~\cite{AHMED2019101}).  Beam from the TRIUMF cyclotron enters through beamline 1V from the left side of the figure and proceeds to the right.  The proton beam is diverted, as required, into beamline 1U by the kicker and septum magnets.  Downstream of the septum, beamline 1U contains bender magnets, quadrupole magnets, and diagnostic elements before ending in the target.  The principal elements relevant to the kicker magnet design are discussed further in the text.
    \label{fig:1VUAschematic}}
\end{figure*}

The layout of beamline 1U is presented in Fig.~\ref{fig:1VUAschematic}.  For more details about the beamline elements, we refer the reader to Ref.~\cite{AHMED2019101}.

Proton beam delivered to beamline 1 comes from extraction port 1, at the south-east side of the cyclotron.  
Magnets steer and focus the beam along the first part, called beamline 1V, towards a tunnel leading into the TRIUMF Meson Hall and housing the kicker magnet.  This location, partially in the cyclotron vault and partially in an alcove, is semi-sheltered from direct radiation from the cyclotron and contains a bender (1BVB2) that splits beamline 1V into beamlines 1A and 1B.
When the kicker is energized, the beam is deflected upward by 12\,mrad.

After the kicker magnet, the beam passes through the de-energized bender 1BVB2 and into a Lambertson septum magnet.
The septum magnet is tilted slightly to remove the ascension in the 1U beam trajectory and to bend the beam 9\degree~to the left, leaving it horizontal and 70\,mm higher than the 1A beam plane.  After passing through a shielding plug, which separates the cyclotron vault from the 1A tunnel, the 1U beam is bent to the left an additional 7\degree~by a small bender magnet (1UB0), through the aperture of additional magnets which provide the desired profile and position on the spallation target.

From the kicker magnet through the septum and beyond, the beam size is of the order of a few millimeters in both {\it x} and {\it y}~\cite{bib:rwang}.
This allows both the kicked (1U) and unkicked (1A) beam to pass through both 1BVB2 and the 48~mm-wide septum magnet constrictions without any significant losses.  The septum between the 1U and 1A beam paths would receive the most beam spill if the kicker magnet was out of synchronization with the notch. After 8 months of running and 1 month of cool-down, the magnet pole material around the septum was measured to be activated to as much as 10~mSv/h on contact. This primarily arose from beam halo.

Several elements in the beamline provide diagnostic data. HARP-style profile monitors (1UHARP0 and 1UHARP2) with 16 wires in a gas-filled chamber are inserted into the beam to accurately measure the beam profile while tuning the beam optics at very low currents, see Section~\ref{sec:DC}. A capacitive probe 1VM4.7 detects the notches between beam pulses and provides timing information for the kicker, see Section~\ref{sec:KSM}. Beam spill monitors (BSM55 and BSM56) give a warning or shut down the beam when they detect high radiation due to the beam hitting the beam tube, see Section~\ref{sec:mistiming}. A toroidal non-intercepting monitor (TNIM) measures the beam current injected into beamline 1U and serves as the main safety device limiting the beam current, see Section~\ref{sec:experiences}.

The beamline ends in a water-cooled neutron spallation target consisting of several tantalum-clad tungsten blocks acting as a beam dump. A prototype UCN source was installed directly above the target in 2017 and first results from the commissioning of the UCN source were reported in Ref.~\cite{UCNprod}.

\section{Design Studies} 

Table~\ref{tab:specifcations} shows the specifications for the kicker magnet system. 
The magnet and power supply were specified based on achieving a deflection angle of 15~mrad.  In the final layout of the beamline a deflection angle of 12~mrad was required. The rise-time specified in Table~\ref{tab:specifcations} (50~$\mu$s) is from 2\% to within $\pm 2$\% of the flat-top field and includes any required settling time for ripple. The kicker system is required to operate in a wide range of repetition rates, from dc mode and single-shot tests during commissioning to up to several hundred Hertz. The initially specified maximum repetition rate was 350~Hz.  For technical reasons related to synchronization with the cyclotron RF, 373~Hz operation of the kicker was needed; hence, the required maximum repetition rate was increased to 400~Hz.

\begin{table}[h!] 
\centering
\caption{Specifications for the UCN kicker magnet system.}
\begin{tabular}{lr}
	\toprule 
\textbf{Parameter} & \textbf{Value} \\ 
	\colrule 
Proton momentum (MeV/c)  & 1090 \\ 
Max. deflection angle (mrad) & 15 \\ 
Magnetic field integral (T$\cdot$m) & 0.0545 \\ 
Aperture, vertical x horizontal (mm$^2$) & 100 x 100  \\ 
Rise time, 2\% to within $\pm 2$\% of flat top ($\mu$s) & 50\\ 
Fall time, 98\% to within $\pm 2$\% ($\mu$s) & 50\\ 
Nominal duration of field flat-top (ms) & 1 \\ 
Continuous repetition rate (Hz) & 0 to 400 \\
Field homogeneity (\%) & $\pm 2.0$ \\
Maximum available length in beamline (m) & 2.0 \\
\botrule
\end{tabular} 
\label{tab:specifcations}
\end{table}

A kicker system was previously designed and built for the ultracold neutron source at Paul Scherrer Institute (PSI), Switzerland, with relaxed specifications compared to our system.  The PSI kicker system is capable of operating at a magnet current of up to 200~A, with a rise time of 500~$\mu$s (5--95\%), and with a 1\% duty cycle (8~s-duration kick every 800~s)~\cite{PSI:Anicic2005}. This kicker magnet had two coils, each with 18 turns. Because of the low repetition rate, average power dissipation in the coils was low, and hence the coils were air-cooled. A four-quadrant converter was used together with both a high-voltage rectifier and a low-voltage rectifier: while ramping up the current, a high voltage of  350~V was applied to the magnet to achieve the 500~$\mu$s rise-time. The low-voltage rectifier, 30~V, was used to maintain the flat-top of the current during the 8~s-long kick. 

Kicker magnets, or chopper dipoles, and power supplies were also required for the 
hadron therapy centers at CNAO~\cite{CNAO:Pullia2005}, Pavia, Italy, and
MedAustron~\cite{medaustron:Benedikt2005,medaustron:Benedikt2010}, Wiener Neustadt, Austria, for turning the beam on and off. The CNAO and MedAustron designs~\cite{Sermeus2004,Dallago2006,Borburgh2010} were chosen as the
baseline for the TRIUMF kicker system because some of their requirements were similar. For example:
\begin{itemize}[leftmargin=*]
\item For the CNAO chopper dipoles a current of 650~A, with
  260~$\mu$s rise and fall times, and a repetition rate of 10~Hz was
  required~\cite{Dallago2006}.
\item For the MedAustron chopper dipoles a current of
  630\,A was required. The specified current rise and fall time was
  250\,$\mu$s, however the magnet insulation was specified such that it
  was capable of 90\,$\mu$s rise and fall time~\cite{Borburgh2010}. A
  maximum repetition rate of 20\,Hz was required.  The current flat-top
  could be from 0\,s to dc~\cite{Stadlbauer2015}, i.e.~no flat-top to continuous flat-top, respectively.
\end{itemize}

The CNAO and MedAustron chopper dipole systems are used to routinely
switch the beam on and off during operation of the medical
facilities~\cite{Bryant2000}, thus the power supplies and magnets are
designed to be reliable~\cite{Kramer2011}.

\begin{figure}[t]
  \includegraphics[width=\linewidth]{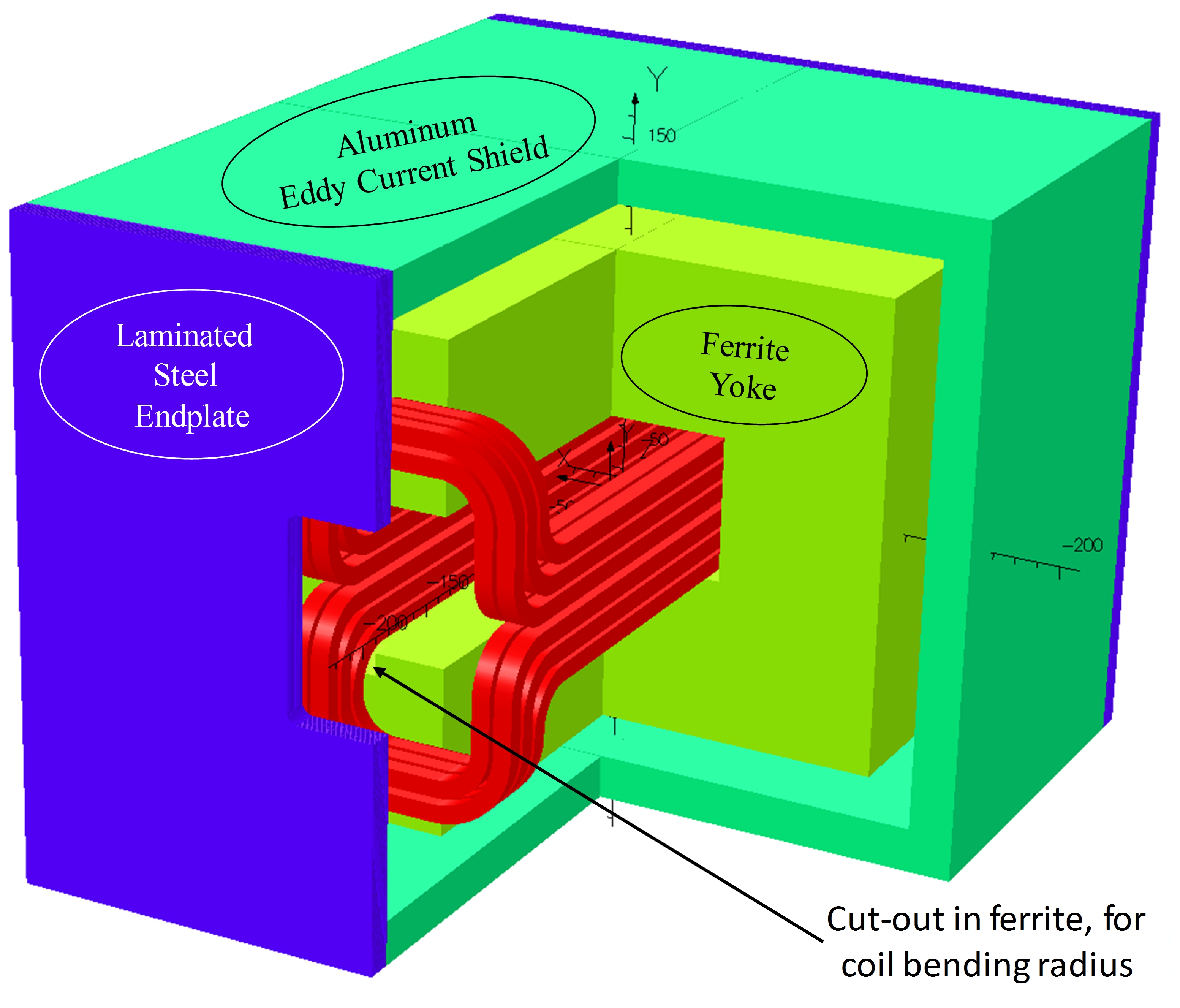}
  \caption{Geometry of MedAustron kicker magnet, which served as a reference design for the TRIUMF kicker magnet. The overall length of the MedAustron magnet, including end plates, is $\sim$0.4~m.}
  \label{fig:medaustron}
\end{figure}

The CNAO chopper dipole uses a window frame construction, named after the shape of its yoke, with
two water-cooled saddle coils of 8 turns each~\cite{Dallago2006}. The
MedAustron chopper dipole (Fig.~\ref{fig:medaustron}) uses a similar construction, but each
saddle coil has 6 turns~\cite{Miro-thesis}. The MedAustron magnet is
housed in a box whose sides, top and bottom are made from 
aluminium. End plates, which act as field clamps, are 12~mm thick and
made out of 1~mm thick steel
laminations~\cite{Kramer2011,Miro-thesis}.  

Although the CNAO and MedAustron systems served as
useful reference designs for the TRIUMF kicker system, the rise and fall times required for the TRIUMF kicker system are a factor of approximately five shorter than those achieved for the CNAO and MedAustron systems. In addition, the repetition rate of the TRIUMF kicker system is at least a factor of 20 above that of the CNAO and MedAustron systems. These factors combined introduce challenges, especially in terms of transient power dissipation in the coils and the power supply voltage required to achieve the specified 50~$\mu$s rise and fall times: the transient power dissipation is significantly increased by skin-effect and proximity-effect in the coils (see section~\ref{sec:FAT}). A key enabling technology in the TRIUMF kicker system is the use of Insulated-Gate Bipolar Transistors
(IGBTs) in the power supply. IGBTs are three-terminal semiconductor devices which combine high efficiency and fast switching and are capable of blocking relatively high voltage and conducting high current. Thanks to careful optimization of
the magnet current, magnet inductance, and power supply
voltage required to meet the challenging specifications for the field rise and fall time and repetition rate, it is not necessary to connect multiple IGBT modules in parallel to achieve the required current, or to connect multiple IGBTs in series to block the required high voltage.

\subsection{Kicker Magnet\label{sec:kickermagnet}}
\begin{figure}[b]
	\includegraphics[width=\linewidth]{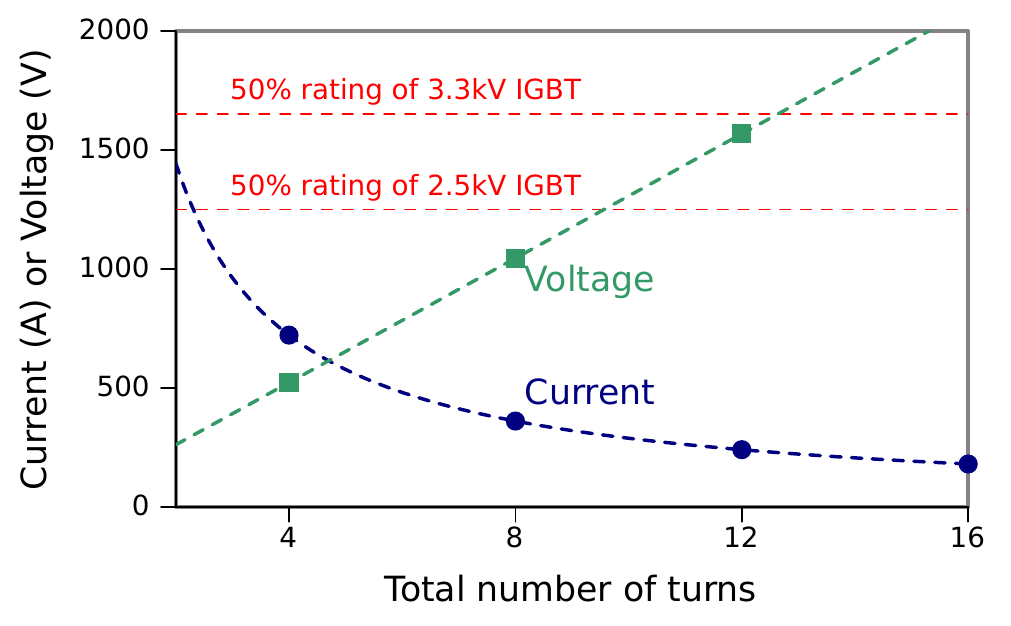}
	\caption{Plot of the required magnet current and estimated magnet voltage to achieve a magnetic field integral of 0.0545~T$\cdot$m, as a function of the total number of turns of the magnet coils with a 1.5~m long yoke. End effects are neglected and $h{\rm(eff)}_{\rm{ap}} \times v_{\rm{ap}} = $ 120~mm $\times$ 100~mm is assumed.}
	\label{fig:VoltageCurrent}
\end{figure}

The available length in the TRIUMF beamline for  the kicker magnet is 2~m. Hence, a mechanical length of 1.5~m was assumed for the magnet yoke to ensure sufficient space for magnetic shielding, electric and water connections, etc. Figure~\ref{fig:VoltageCurrent} shows a plot of the required magnet current, as a function of the total number of turns of the kicker magnet coils, to achieve a magnetic field integral of 0.0545~T$\cdot$m. Assuming that the yoke is unsaturated and has a high relative permeability, the required current ($I$) is calculated from~\cite{Barnes:CAS2017}:
\begin{equation}\label{Ieqn}
I = \frac{B_x v_{\rm{ap}}} {\mu_{\rm{0}} N},
\end{equation}
where $B_x$ is the nominal  flux density (36.3~mT, for a yoke with a length of 1.5~m parallel to the direction ($z$) of propagation of the beam) in the aperture, $N$ is the number of turns, $v_{\rm{ap}}$ is the distance between the legs of the yoke
and $\mu_{\rm{0}}$ is the permeability of free space. 

The design of the UCN kicker magnet is based on those of CNAO and MedAustron (Fig.~\ref{fig:medaustron}). 
It also uses a window frame construction with two, nominally identical, water-cooled saddle-shaped coils. 
In addition, each saddle-shaped coil has a double-layer pancake arrangement: hence the total number of turns is an integer multiple of four. From Fig.~\ref{fig:VoltageCurrent} a total of 4 turns require a current of approximately 720~A to achieve a magnetic field integral of 0.0545~T$\cdot$m: this current is reduced to 360~A, 240~A and 180~A for a total number of turns of 8, 12 and 16, respectively. In order to determine the voltage required to increase the current from 0~A to the required flat-top value, or to decrease the current from the required flat-top value to 0~A, in 50~$\mu$s, it is necessary to estimate the inductance of the magnet. 

Neglecting end effects,  the inductance of the magnet is calculated from: 
\begin{equation}\label{Leqn}
L = \frac{\mu_{\rm{0}} N^2 h_{\rm{ap}} l}{v_{\rm{ap}}}
\end{equation}
where $l$ is the length of the yoke. For a fast single-turn kicker magnet,  $h_{\rm ap}$ is the distance between the inside edges of the conductors~\cite{Barnes:CAS2017}. However, each saddle coil of the UCN kicker magnet has two layers and each layer conducts equal currents. Thus, the effective value $h{\rm(eff)}_{\rm ap}$ is dependent upon the dimensions of the conductors and the distance between the two layers. As a first approximation, for a 100~mm aperture and assuming 8~mm thick conductors, an effective value, for $h{\rm(eff)}_{\rm ap}$, of 120~mm was assumed. Equation~(\ref{Leqn}) shows that the magnet inductance increases in proportion to the square of the number of turns, however the current required decreases in proportion to the number of turns (equation~(\ref{Ieqn})). Hence, to achieve 50~$\mu$s rise and fall times, the required magnet voltage increases linearly with the number of turns: this is shown in Fig.~\ref{fig:VoltageCurrent}. The power supply output voltage will be higher than the magnet voltage due to inductance of the cables connecting the magnet and supply. Hence, to minimize the required power supply voltage, the inductance of the supply and cables must be kept to a reasonable minimum. 

IGBTs are used for the active power semiconductor devices in the kicker magnet power supply. To ensure long-term  reliability of the IGBTs, it is preferable that they are used at no more than 50\% of their rated voltage. The 50\% de-rating is chosen to avoid failure due to, for example, cosmic radiation and neutrons~\cite{Shoji}. Assuming a single IGBT is used, Fig.~\ref{fig:VoltageCurrent} shows lines for 50\% of the rated voltage for 2.5~kV and 3.3~kV rated IGBTs, which were readily available ratings at the time of the initial system design. 

\begin{figure}[t]
	\includegraphics[width=\linewidth]{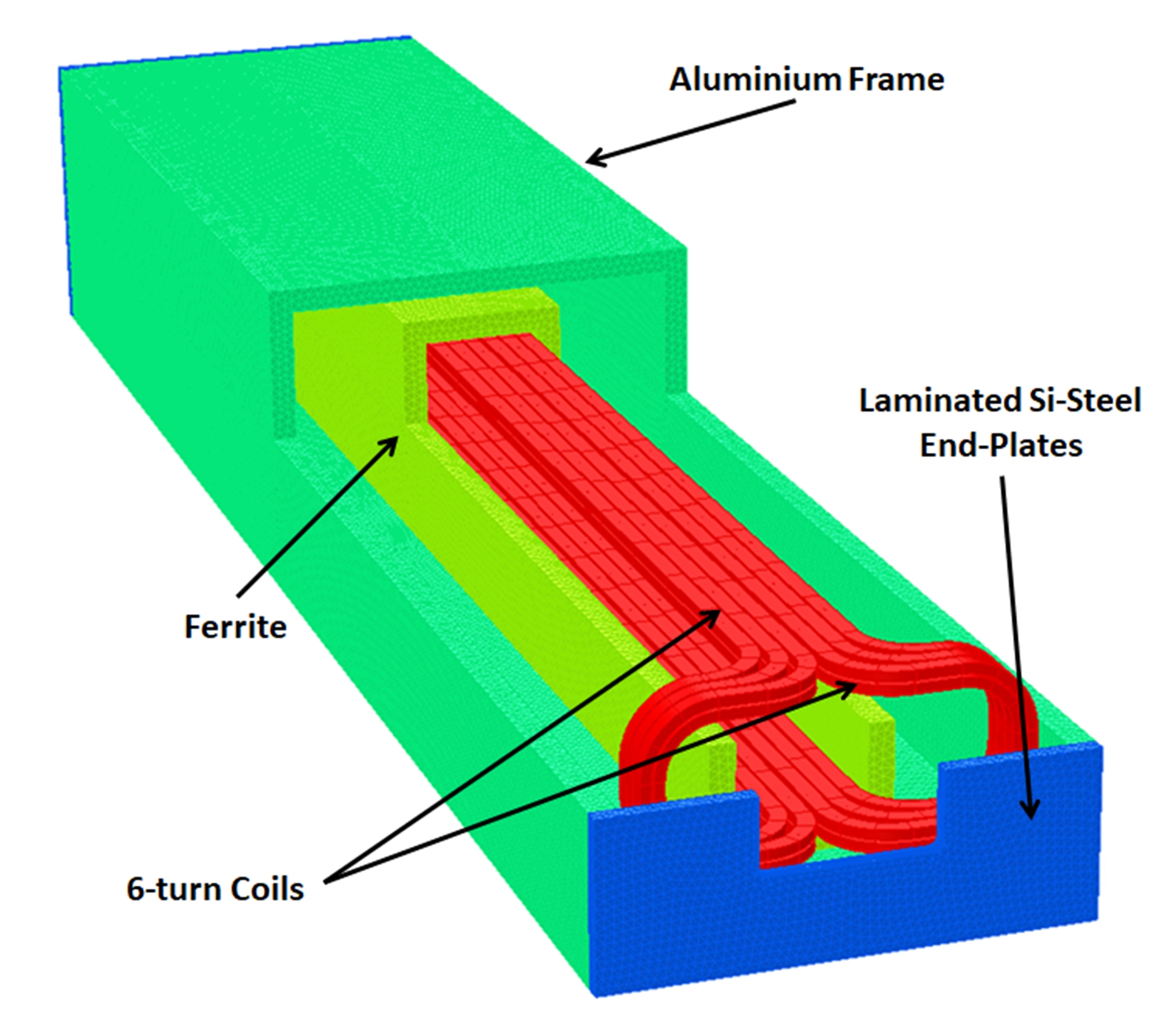}
	\caption{Geometry for a 3D simulation of the UCN kicker magnet. The overall length of the magnet, including end plates, is $\sim$1.71~m.}
	\label{fig:UCN-magnet}
\end{figure}
It is desirable that the UCN kicker system can operate in a dc on mode. The dc resistance of the magnet coil is proportional to the total length of the coils: to a first approximation the length, and hence dc resistance, is proportional to the total number of turns. The dc conduction losses in the magnet coil are given by the dc resistance multiplied by the square of the current. For a representative dc resistance of 1\,m$\Omega$ per turn, the dc conduction losses for the coil would be 2090\,W, 1050\,W, 700\,W and 520\,W for 4, 8, 12 and 16 turns, respectively. Thus, to reduce dc conduction losses, the current required is decreased by increasing the number of turns, however a higher voltage is required to achieve the specified rise and fall times. Based on considerations of the current required  to achieve a magnetic field integral of 0.0545\,T$\cdot$m, the voltage needed to obtain 50\,$\mu$s rise and fall times, and the dc power loss in the coil, a total of 12 turns was chosen for the UCN kicker magnet. The inductance of the 12-turn kicker magnet, estimated from equation~(\ref{Leqn}), assuming $h{\rm(eff)}_{\rm{ap}} \times v_{\rm{ap}} = 120$\,mm$\,\times 100$\,mm and neglecting end-effects, is $\sim$326\,$\mu$H for the 1.5\,m yoke length. 
To obtain a deflection of 15\,mrad with a 12-turn coil requires a current of approximately 240\,A. For a magnet inductance of 326\,$\mu$H, to achieve a current rise-time of 50\,$\mu$s, a magnet voltage of 1.56\,kV is required.

The ferrite yoke and the saddle coils are enclosed by a 20~mm thick aluminium frame and laminated steel end plates (Fig.~\ref{fig:UCN-magnet}). The aluminium and the steel are used as eddy-current screens to limit radiation of electromagnetic noise. The laminated steel end plates also reduce fringe fields at both ends of the kicker magnet. Each end plate is 10~mm thick and constructed from 0.35~mm-thick laminations. The Opera~\cite{cobham} software has been used to carry out detailed electromagnetic simulations of the kicker magnet~\cite{Hahn}. These simulations have permitted both the saddle coils and the ferrite yoke dimensions to be optimized to achieve the required field uniformity, limit the power dissipation in the coils, and to select the cross-section of the ferrite yoke. The build-up of the ferrite yoke, in the directions perpendicular to the beam direction, is 20~mm.

\begin{figure}[t]
	\includegraphics[width=\linewidth]{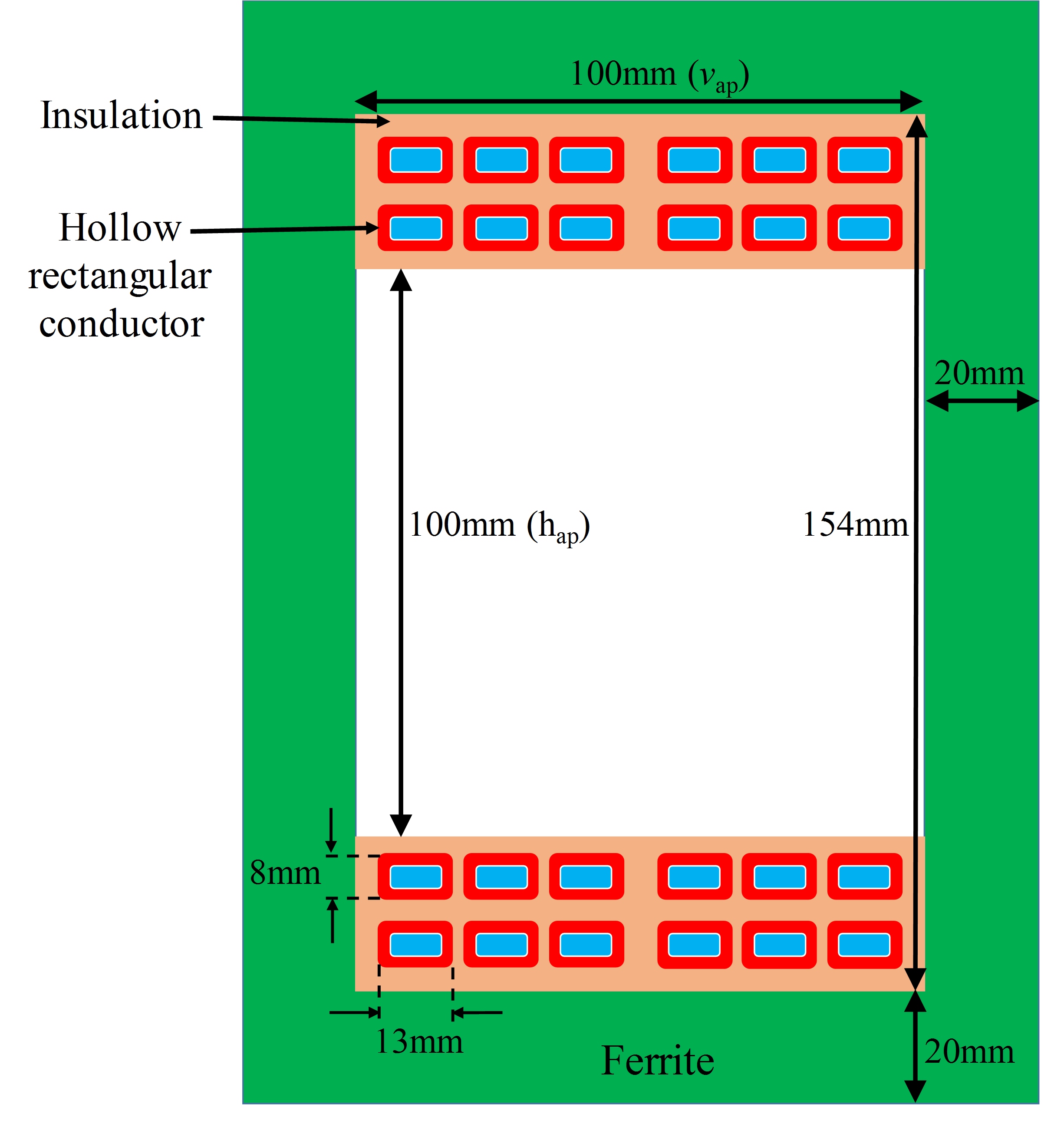}
	\caption{Cross-section of aperture of kicker magnet: rectangular coil conductors with 2~mm wall (red), insulation (brown), ferrite (green) and 100~mm $\times$ 100~mm aperture (white).}
	\label{fig:ApertureCrossSection}
\end{figure}

The 3D model was used to calculate the uniformity of the deflection field, by integrating the predicted magnetic field along lines parallel to the beam direction, through the aperture of the kicker magnet~\cite{Hahn}. The field in the central 80~mm $\times$ 80~mm region of the aperture was studied as a function of the vertical spacing between adjacent conductors of the coil. The best homogeneity ($\pm 1$\%) is achieved with the conductors within each saddle coil equally distributed  (Fig.~\ref{fig:ApertureCrossSection})~\cite{Hahn}.  The spacing between each coil allows for electrical insulation. 

\begin{figure}[t]
	\includegraphics[width=\linewidth]{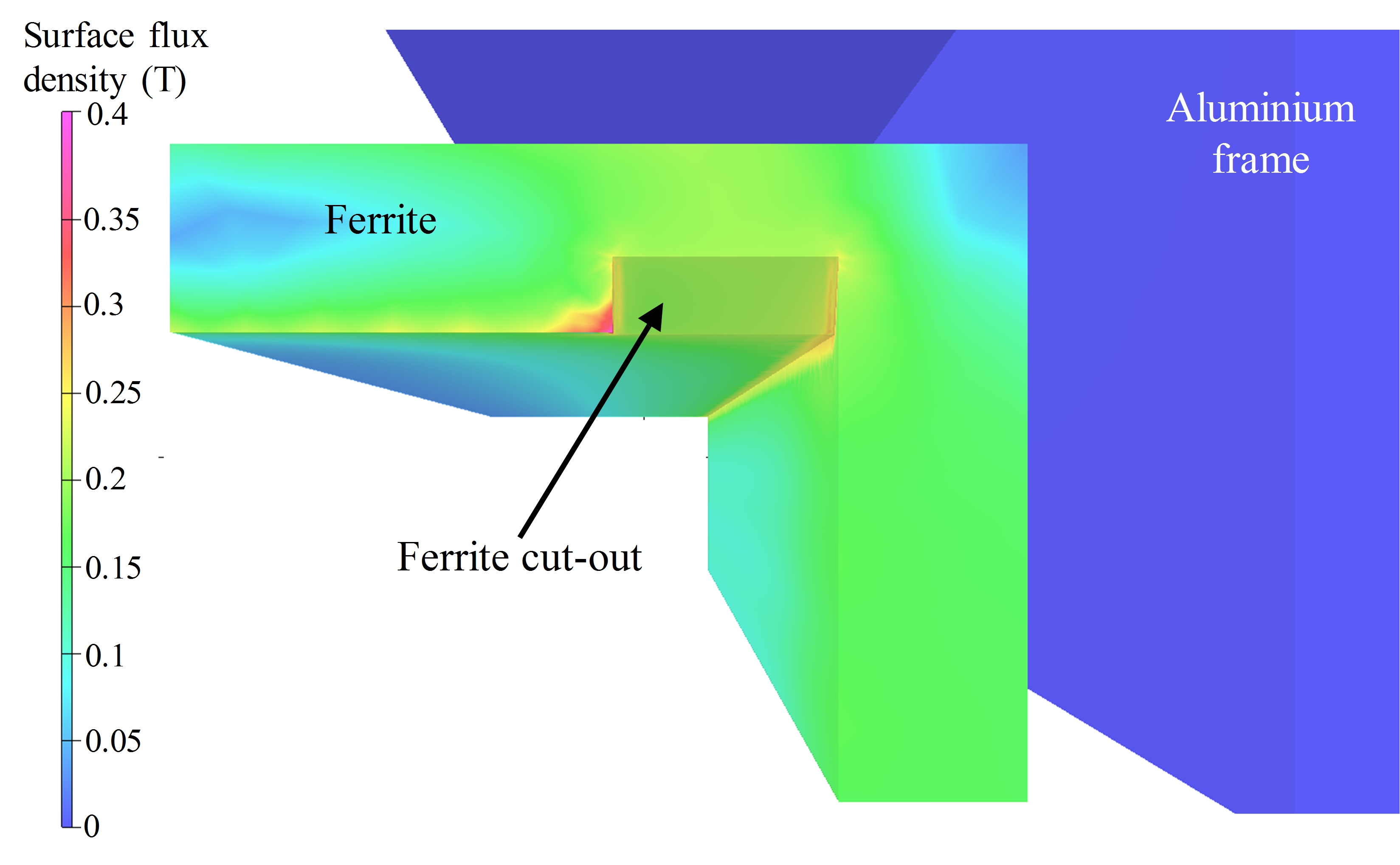}
	\caption{Initial design of end ferrites, with cut-out to bend the saddle coils towards the aluminium frame (quarter of geometry shown). The colour regions  associated with the ferrite represent surface contours of predicted flux density, for a coil current of 240~A.}
	\label{fig:FerriteCutOut}
\end{figure}

As mentioned above, the design of the UCN kicker magnet is based on that of both MedAustron and CNAO chopper dipoles: Fig.~\ref{fig:medaustron} shows the MedAustron chopper dipole with a cut-out in the ferrite yoke, to allow for the coil bending radius, to limit the overall length of the magnet. Hence, the initial design of the end ferrites of the UCN kicker magnet also had a cut-out machined in the ferrite yoke (Fig.~\ref{fig:FerriteCutOut}) so that the saddle shaped conductors could be bent towards the aluminium frame with the required bending radius and without extending the length of the coils. Fig.~\ref{fig:FerriteCutOut} also shows contours of the predicted flux density on the surface of the ferrite, for a coil current of 240~A: the highest flux density is close to the edges of the cut-out. 

Machining a cut-out in the ferrite increases difficulty of manufacture and therefore the cost of the ferrite. Hence,  
Opera3D simulations were carried out to assess the influence of increasing the length of the saddle coils, for a given yoke length, to remove the machined cut-outs from the ends of the UCN magnet ferrite yoke---the resulting geometry is shown in Fig.~\ref{fig:UCN-magnet}. The simulations show that increasing the overall length of each saddle coil by 38~mm, to remove the cut-outs from the ends of the ferrite yoke, further improved the homogeneity of the integrated field from $\pm 1$\% to $\pm 0.8$\%~\cite{Hahn}. The specified integrated field uniformity is $\pm 2$\% (Table~\ref{tab:specifcations}), thus the predicted field uniformity meets the specifications. The effective magnetic length is predicted to be 1.60~m. The overall mechanical length of the kicker magnet, including the laminated steel plates at each end, is $\sim$1.71~m. 

Figure~\ref{fig:ApertureCrossSection} shows a cross-section of the coil conductors in the aperture of the kicker magnet. 
 From Fig.~\ref{fig:ApertureCrossSection}, for the optimized distribution of the conductors and allowing for 2~mm of insulation around each conductor and an extra 1~mm around each saddle coil, the effective value of $h_{\rm{ap}}$ is the distance between the mid-point of the layers of the saddle coils, i.e.~126~mm. Substituting $h{\rm(eff)}_{ap} = $ 126~mm in equation~(\ref{Leqn}) gives an inductance of $\sim$342~$\mu$H for the 1.5~m yoke length.  The corresponding inductance calculated from Opera3D dc simulations is $\sim$360~$\mu$H. However, this is considered an overestimate of the value: in the 3D dc simulations the coil is modelled as being solid, i.e.~no water cooling channel (see below), and current is considered to be evenly distributed throughout each turn of the coil. Hence, in the 3D dc model, field can pass through the coils, resulting in an increase of the predicted stored energy and inductance. 

A hollow rectangular conductor with external dimensions of  13~mm $\times$ 8~mm and a uniform wall thickness of 2~mm (\#8905 from Luvata~\cite{Luvata}) was chosen for the saddle coils (Fig.~\ref{fig:ApertureCrossSection}): the 9~mm x 4~mm internal channel is  for water cooling. The two saddle shaped coils are electrically in series but water cooled in parallel. The power dissipation was computed with realistic trapezoidal current waveforms, using the Opera2D transient solver. As a result of eddy current and proximity effects the total resistance of the two series coils changes considerably during the current pulse; the resistance of the coil reduces by an order of magnitude from the start of the current flat-top to the end of a 1~ms current flat-top. In addition, as a result of the decay of eddy currents in the coil the field penetrates further into the coil (see Figs.~22 and 23 of \cite{Barnes:CAS2017}) and, hence, the internal inductance of the coils increases: as a result the total inductance of the two series coils increases by several percent during the flat-top of the current pulse. The rms value  for a trapezoidal current waveform with 50~$\mu$s rise and fall times and 240~A flat-top current for 1~ms is $\sim$147~A at a repetition rate of 350~Hz and $\sim$157~A at 400~Hz. The power loss predicted by Opera2D for 350~Hz operation, taking into account eddy-currents, is 775~W for the 12 turns, which corresponds to an equivalent series resistance of $\sim$37~m$\Omega$, i.e.~a factor of three higher than the dc resistance. The power loss of 775~W is 18\% greater than the power loss for 240~A dc. The two saddle coils are water cooled in parallel. A water pressure of 4~bar gives a flow of approximately 5.5~l/min per saddle coil. The calculated change in water temperature, with 775~W of dissipation, is approximately 1$^\circ$C. 

The kicker magnet is designed to be outside vacuum. Hence a vacuum tube is required in the aperture of the magnet, which must provide a good vacuum connection to the adjacent beampipes. This vacuum tube is not a part of the commercial order for the kicker magnet and was added at TRIUMF. A metallic beampipe,  in the aperture of a kicker magnet, should be avoided: during field rise/fall the changing field will result in 
eddy currents which create a reaction field that opposes the changing magnetic field. Eddy currents in the metallic beampipe would transiently shield the beam from the changing magnetic field of the kicker magnet,  increasing field rise and fall times. In addition, the inductance of the magnet would transiently be reduced,  increasing the rate-of-rise of current of the power supply. Therefore, alumina is frequently used for the tube in a kicker magnet~\cite{PSI:Anicic2005, PAC07-Barnes, He2002, IPAC11-Pont, APAC01-Ghodke, IPAC12-Barnes}. However, a thin metallic coating is frequently applied to the inner surface of the alumina tube to both screen the kicker magnet yoke from the beam~\cite{He2002, IPAC11-Pont, APAC01-Ghodke, IPAC12-Barnes}, and also to prevent build-up of static charge on the tube~\cite{APAC01-Ghodke}. 

The solution adopted at TRIUMF for the vacuum tube was a non-conducting, uncoated, tube: this is discussed further in Section~\ref{sec:prebeam}.

\subsection{Power Supply}

\begin{figure}[t]
	\includegraphics[width=\linewidth]{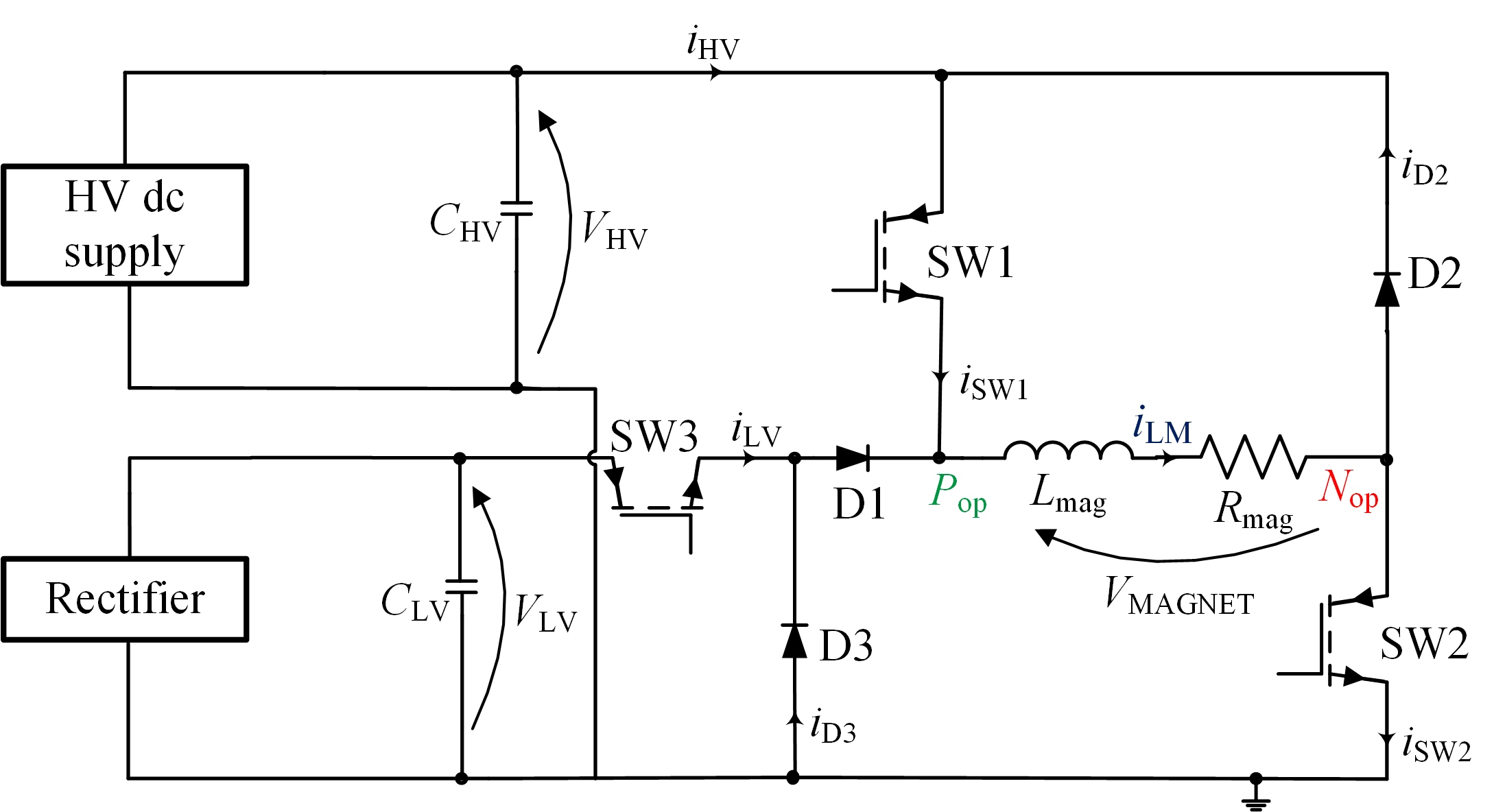}
	\caption{Simplified electrical schematic of the UCN kicker system (described in the text).}
	\label{fig:powersupply}
\end{figure}

The pulse power modulator (Fig.~\ref{fig:powersupply}) consists of a high-voltage (HV) stage, for generating the rising and falling edges of the output current pulse, and a low-voltage stage for maintaining the flat-top of the output current pulse. The high-voltage stage uses an HV dc supply that charges capacitor $C_{\rm HV}$. The low voltage stage uses three-phase rectifiers supplied by the secondary windings of a dedicated transformer. The kicker magnet and the parasitic inductance of the cable is represented by inductor $L_{\rm mag}$ together with a series resistor, $R_{\rm mag}$. The quantity $R_{\rm mag}$ accounts for both the parasitic resistance of the magnets and the resistance of the cables (see below). There are three power IGBTs, labelled SW1, SW2 and SW3, for switching on and off the magnet current: SW1 and SW2 switch high voltage ($V_{\rm HV}$). In addition, there are three power diodes, labelled D1, D2 and D3: diodes D1 and D3 provide a free-wheeling path for magnet current, when SW1 is turned off. IGBT SW3 switches at relatively high frequency (few kHz) to regulate the flat-top of the magnet current (see below). Hence, diode D1 blocks high voltage, so that SW3 can be rated at relatively low voltage.

The operation of the circuit can be divided in to the following phases:
\begin{itemize}[leftmargin=*]
	\item	The circuit is in the OFF state, all the switches are off, and no current is flowing in the magnet.
	\item	A control signal arrives, before a START trigger and after a previous STOP trigger, to program the magnitude of magnet current to be used during the next cycle by the power supply current control. The high voltage is set in proportion to the set current magnitude. Hence, the current rise and fall times  (see below) are almost constant over the operating range of the system.
	\item	When the START command is received the control circuitry turns both SW1 and SW2 on. In this way the output voltage of the high voltage stage is applied to the kicker magnet and a current starts to flow with a positive sign, given the reference direction for the magnet current (${i_{\rm{LM}}}$). The current increases as an exponential with a time constant approximately equal to (${L_{\rm{mag}}}/({R_{\rm{mag}}}+{R_{s}}$)), where ${R_{s}}$ is the sum of the  on-state resistances of switches SW1 and SW2. However, the rising current can be considered to be a linear ramp because the steady-state value is much higher than the required maximum current. The value of capacitor ${C_{\rm{HV}}}$ is chosen to provide the charge required for the rising current, with an acceptable reduction in voltage (approximately 1.5\%) across the high voltage capacitors.

	\item	After a time interval, $t_{\rm{r}}$, approximately equal to  $(({L_{\rm{mag}}}\cdot {i_{\rm{LM}}})/ {V_{\rm{HV}}})$ the required magnet current is reached. At this point switch SW1 is turned off. The magnet current starts to freewheel through diode D3, diode D1 and SW2. From this instant on the flat-top current is maintained by switching SW3 on and off. Switch SW3 connects the output of the low-voltage stage of the power supply to the magnet. Hysteresis control is used for SW3: as soon as the actual value of the magnet current is greater than a certain upper threshold SW3 is switched off, and then back on when the magnet current reduces below a certain lower threshold. This flat-top phase can last indefinitely, however the nominal duration is approximately 0.9~ms.

	\item	When the STOP command is received both SW2 and SW3 (if SW3 was on) are switched off simultaneously. The magnet current is then diverted to the capacitor $C_{\rm{HV}}$ through diodes D3, D1 and D2. Since the voltage applied to the magnet has a similar  magnitude but opposite polarity with respect to the rising ramp, the current fall time is almost equal to the current rise time. When the magnet current reaches 0~A the diodes naturally turn off and the circuit is again in the OFF-state.
\end{itemize}

The safe state for the TRIUMF beam line is when zero current is delivered to the UCN kicker magnet. Hence precautions are taken to avoid a failure in the switching power stages of the modulator.  This protection is achieved by choosing an IGBT, the FD500R65KE3-K from Infineon~\cite{Infineon}, rated at 6.5~kV for SW1 and SW2, i.e.~a factor of three to four above the operational voltage  of the high voltage stage. The IGBT chosen is suitable for chopper applications and traction drives, and is rated for 500~A continuous dc collector current, i.e.~a factor of more than two above the maximum dc current rating required for 15~mrad deflection with the 12-turn coil (240~A). In addition, the IGBT package has an AlSiC base plate for increased thermal cycling capability. The FD500R65KE3-K IGBT modules contain a diode rated at 6.5~kV and 500~A: this diode in the IGBT module for SW2 is used for D2 and the diode in the IGBT module for SW1 is used for D1 (Fig.~\ref{fig:powersupply}). The IGBT module used for SW3 is the FD1400R12IP4D from Infineon~\cite{InfineonSW3}, rated at 1.2~kV and 1400~A: this module, which also contains a diode, is typically used in chopper applications.
In case of failure of an IGBT, the magnet current can be reduced rapidly to 0~A. This is achieved by opening a contactor in series with each of the outputs of the power supply  
(not shown in Fig.~\ref{fig:powersupply}). Metal Oxide Varistors connected to ground, on   the magnet side of each contactor, absorb the magnet energy if the contactor opens when load current is flowing in the magnet.

The power supply and magnet are connected by a length of $\sim$17~m of cable: the power supply is on the TRIUMF cyclotron vault roof, in a non-radiation area. Since the UCN kicker system is required to pulse at a frequency of up to 400~Hz, with a relatively fast current ramp rate,  generation of electromagnetic noise was a real concern. Hence special care was taken choosing the cable. Coaxial cables were considered but were discounted for two reasons:
\begin{itemize}[leftmargin=*]
	\item To use the coaxial cable as intended, current should flow through the central conductor and return through the outer conductor, i.e.~the central conductor should be connected to one end of the kicker magnet and the outer conductor to the other end. However, during the rising edge of the current pulse, one end of the magnet is at high-voltage and during the falling edge the opposite end of the magnet is at high voltage, i.e.~the outer conductor of the coaxial cable would be at high voltage either during the rising or falling edge of the magnet current. This effect could be a significant source of radiated noise. An alternative would be to use two separate coaxial cables, with the central conductor of one cable connected at one end of the magnet and the central conductor of the other connected to the opposite end of the magnet. Even though the outer conductors of the cables could be connected to one another, the go and return currents are no longer flowing coaxially, but rather in two separate cables.
	\item The magnet and power supply should be able to operate in dc on mode: the resulting rms current is relatively high and several parallel coaxial cables would be required, resulting in a bulky and relatively expensive arrangement.
\end{itemize}
Instead, a three-phase power cable (Okonite CLX 571-23-3244~\cite{Okonite}) was chosen. This type of cable is frequently used in variable speed drive applications, where there are rapidly changing voltages present. The phase conductors are stranded copper, and are thus ideal for an application such as the UCN kicker system. The cable chosen was available from the manufacturer's stock and relatively inexpensive. This cable is rated at 15~kV; the current rating is 525~A at 40$^\circ$C, in a cable tray, hence there is a significant safety margin in comparison with the operational duty of the UCN kicker system. Two of the three-phase conductors are used, one as a go and the other as a return conductor. The aluminium sheath, which is continuous and hence is a good electromagnetic shield, is connected to ground at the power supply end and connected to the magnet frame at the magnet end. The unused phase conductor is, for safety reasons, connected to ground at the power supply end and unconnected at the magnet end.   The measured characteristic impedance of Okonite CLX 571-23-3244, between the two phase conductors used, is approximately 20~$\Omega$, which corresponds to an inductance of $\sim$167~nH/m and a capacitance of $\sim$420~pF/m; the 167~nH/m is derived from short-circuit tests carried out during factory acceptance tests (FATs) - see below. 

Following the ramp-down of the current pulse, once the power semiconductor switches and diodes on the output of the modulator turn-off, the cables charge to a voltage whose magnitude is dependent upon the off-state impedance of the power switches and diodes. Charging of the cable excites an oscillation whose frequency (12.6~kHz) is determined mainly by the inductance of the kicker magnet ($\sim 320~\mu$H - see below) and the capacitance of the cable: this can result in an oscillatory post-pulse current in the magnet unless adequate precautions are taken. The oscillatory current could deflect beam that should be undeflected; the magnitude of undesired deflection would be dependent upon the magnitude of the oscillatory current. To damp this oscillation, a filter, consisting of a capacitor and series resistor, is installed at each end of both used phase conductors of the cable. At the power supply end of each cable the filter is connected between the conductor of the cable and ground. At the magnet end the filter is connected between the phase conductor of the cable and the cable sheath.

\section{Performance Tests Before Beam Delivery}
\subsection{Factory Acceptance Tests}
\label{sec:FAT}
\begin{figure}
	\includegraphics[width=\linewidth]{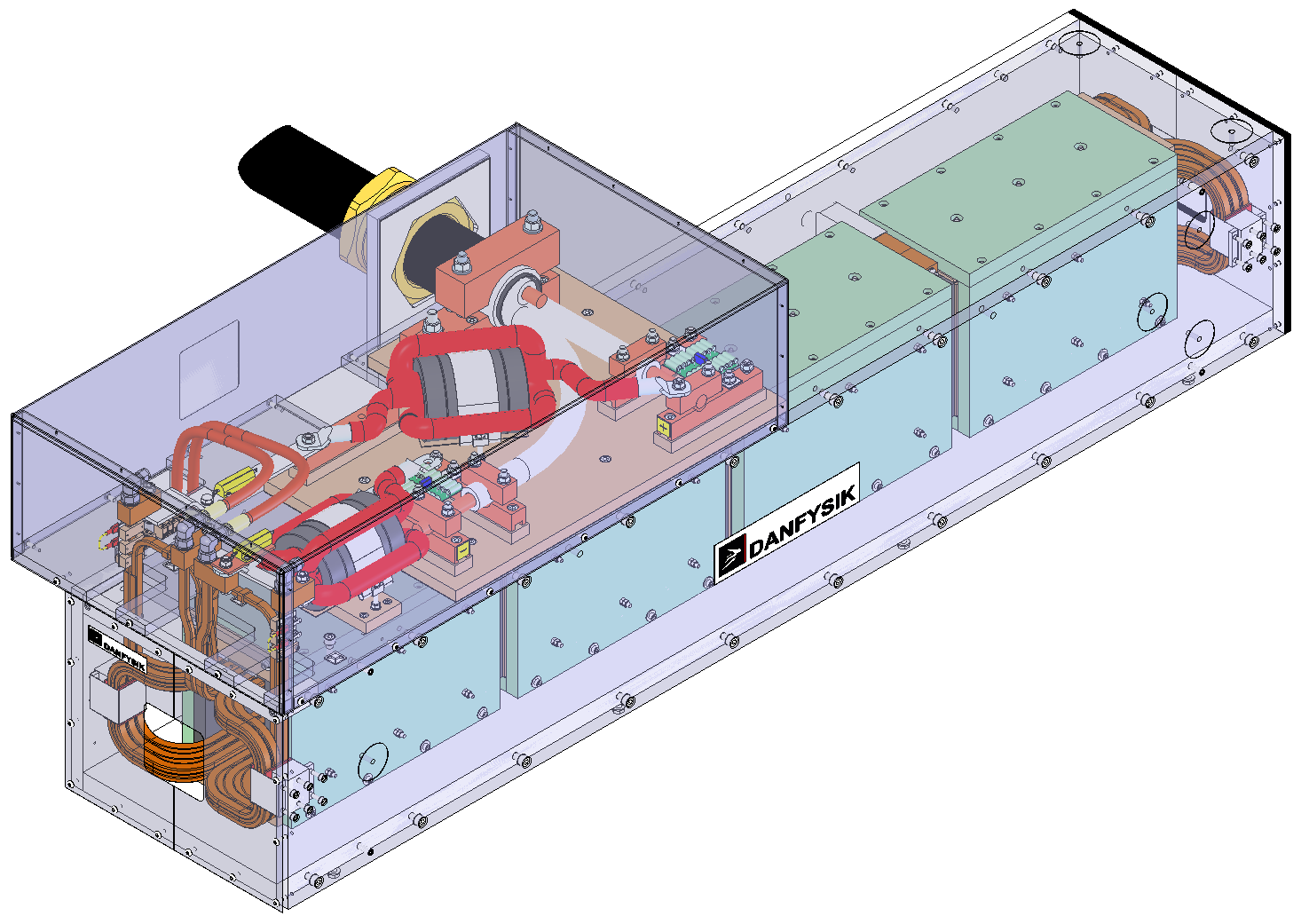}
	\caption{3D engineering drawing of the UCN kicker magnet and connection box. The overall length of the magnet is $\sim$1.71~m.}
	\label{fig:drawing}
\end{figure}

Danfysik~\cite{Danfysik} won the contracts to build both the UCN kicker magnet and power supply. Figure~\ref{fig:drawing} shows a 3D engineering drawing of the kicker magnet.
Several tests and measurements were specified to be carried out during the FATs of the power supply and magnet, including:
\begin{itemize}[leftmargin=*]
	\item Measurement of magnet inductance, coil resistance, and capacitance between coil and aluminium frame.
	\item Field mapping, using a Hall probe, at a current of 243~A dc.
	\item Measurement of current waveform.
    \item Power supply operation with the kicker magnet replaced by a short-circuit.
\end{itemize}

The measured inductance of the UCN kicker magnet is 328.2~$\mu$H at 1~kHz and 321.8~$\mu$H at 7~kHz: the 7~kHz corresponds approximately to the bandwidth of a pulse with 50~$\mu$s rise and fall times~\cite{Tektronix:XYZs}. The dc resistance of the 12 turns is 12.1~m$\Omega$ at 21.8~$^\circ$C. The capacitance, measured between one terminal of the coil and the aluminium frame, is 795~pF.

A Hall probe was used to map the field from the longitudinal centre of the magnet to a distance of 225\,mm outside each end of the laminated steel end plates. This was repeated such that the magnet aperture was scanned horizontally every 10\,mm and vertically every 10\,mm to obtain the field map on an area of $\pm 40$\,mm  by $\pm 40$\,mm. The measured magnetic field at the centre of the aperture was 36.37\,mT, which is very close to the expected value of 36.30\,mT. The effective magnetic length, along the centre line of the aperture, is 1574\,mm; the error is 1.6\% with respect to the 1600~mm predicted using Opera3D. The lower effective length can be compensated by slightly increasing the magnet current.
The homogeneity of the integrated-field, derived from the measurements, was $\pm 1.4$\%, compared to $\pm 0.8$\% predicted. Nevertheless, the specified integrated-field uniformity is $\pm 2$\% (Table~\ref{tab:specifcations}); thus the measured field uniformity meets the specifications. 

\begin{figure}[t]
	\includegraphics[width=\linewidth]{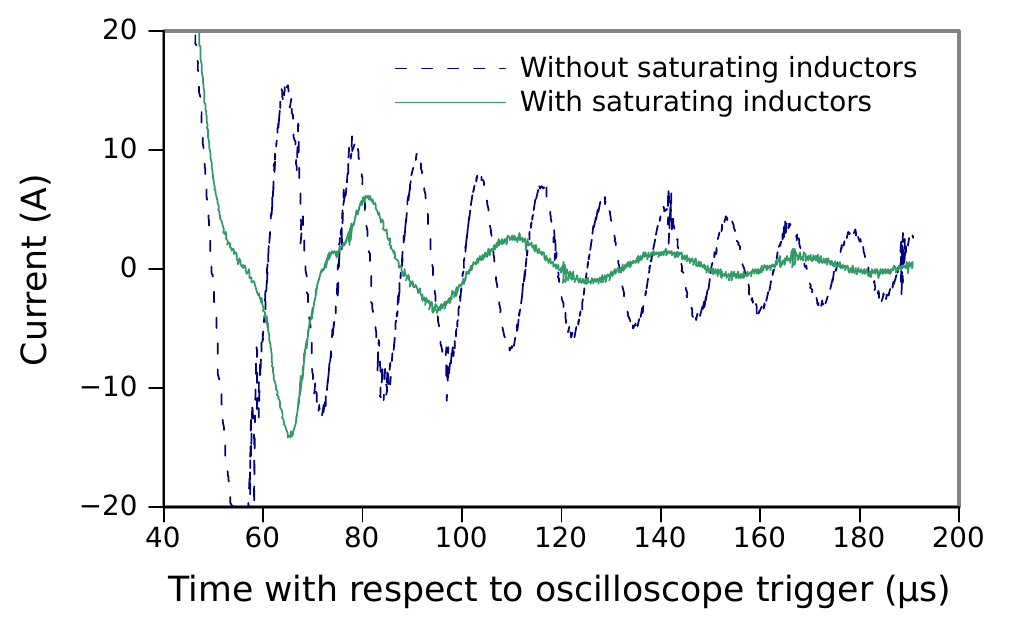} 
	\caption{Measured post-pulse current following 250~A flat-top magnet current: without saturating inductors (blue dashed trace) and with saturating inductors (green solid trace).}
	\label{fig:initialundershoot}
\end{figure}

\begin{figure}[h]
	\includegraphics[width=\linewidth]{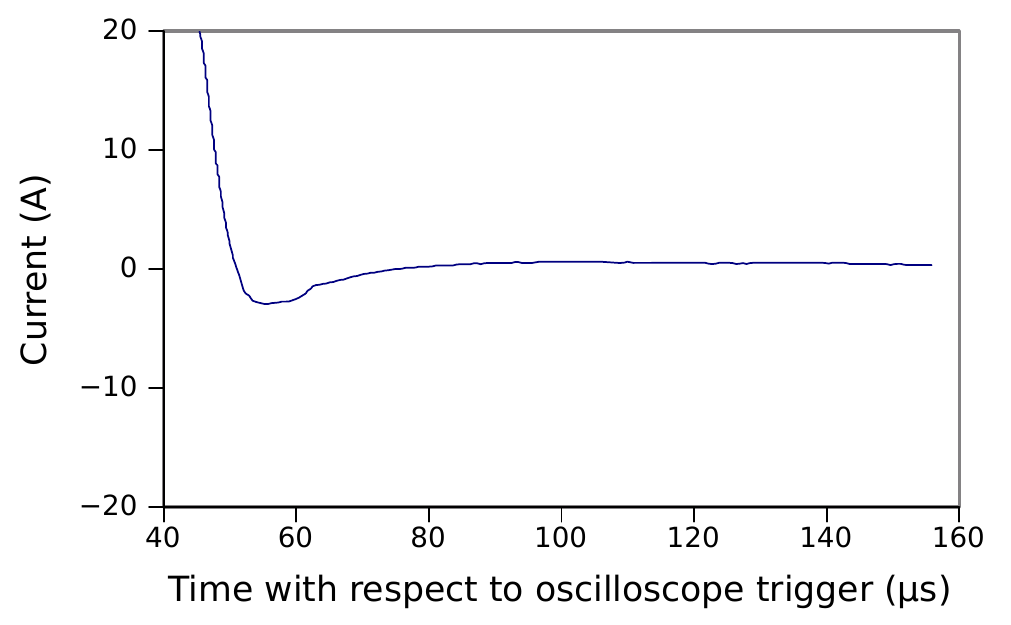} 
	\caption{Measured post-pulse current following 250~A flat-top magnet current, without saturating inductors but with SF5408 fast recovery diodes, in a series/parallel arrangement, connected in series with diode D2.}
	\label{fig:finalundershoot}
\end{figure}

The power supply and kicker magnet at TRIUMF are connected by a 17~m long cable. However, at Danfysik, 30~m of cable were used for the FATs, as this length was already available on a cable drum. The filters installed at both the power supply and magnet end of the cable were expected to result in relatively rapid damping of any  oscillations excited due to charging of the cable at the bottom of the falling edge of the current pulse. However, during the FATs, it was observed that the reverse recovery charge of diode D2  (Fig.~\ref{fig:powersupply}), which was neglected in simulations, results in a short period of reverse current in the kicker magnet at the end of current ramp-down. When this current ``snaps off" it exacerbates the oscillatory behaviour: during the FATs  the peak magnitude of this oscillation was measured to be 22~A (blue dashed trace in Fig.~\ref{fig:initialundershoot}) for a flat-top magnet current of 250~A, which was unacceptably high. To reduce the peak negative current two significant changes were made during the FATs:

\begin{itemize}[leftmargin=*]
	\item Saturating inductors were installed at the magnet end of both used Okonite CLX phase conductors. They provide high inductance at low current, greatly reducing the rate of change of current at low current. A resistor in parallel with the saturating inductor enhances damping. The resulting current measurement (green solid trace in Fig.~\ref{fig:initialundershoot}) shows that the saturating inductors decrease the peak  magnitude of the oscillation from 22~A to 14~A. Additionally, the frequency of the oscillation is reduced from approximately 80~kHz to 30~kHz. The saturating inductors, together with their parallel connected resistors, result in relatively rapid damping of the post-pulse oscillation.
	\item Fast recovery diodes (SF5408) with very low reverse recovery charge, in a series/parallel arrangement, were connected in series with diode D2. For a flat-top current of 250\,A the fast recovery diodes result in a reduction of the peak magnitude of the reverse current in the magnet to 3\,A. The resistor in parallel with each saturating inductor, together with a filter consisting of 2.2\,nF in series with 47\,$\Omega$, at each end of both used phase conductors, is very effective at damping oscillations. Figure~\ref{fig:finalundershoot} shows the resulting post pulse current.  The undershoot, 1.2\%, is well within the specification of $\pm 2$\% (Table~\ref{tab:specifcations}).
\end{itemize}

\begin{figure}[h]
	\includegraphics[width=\linewidth]{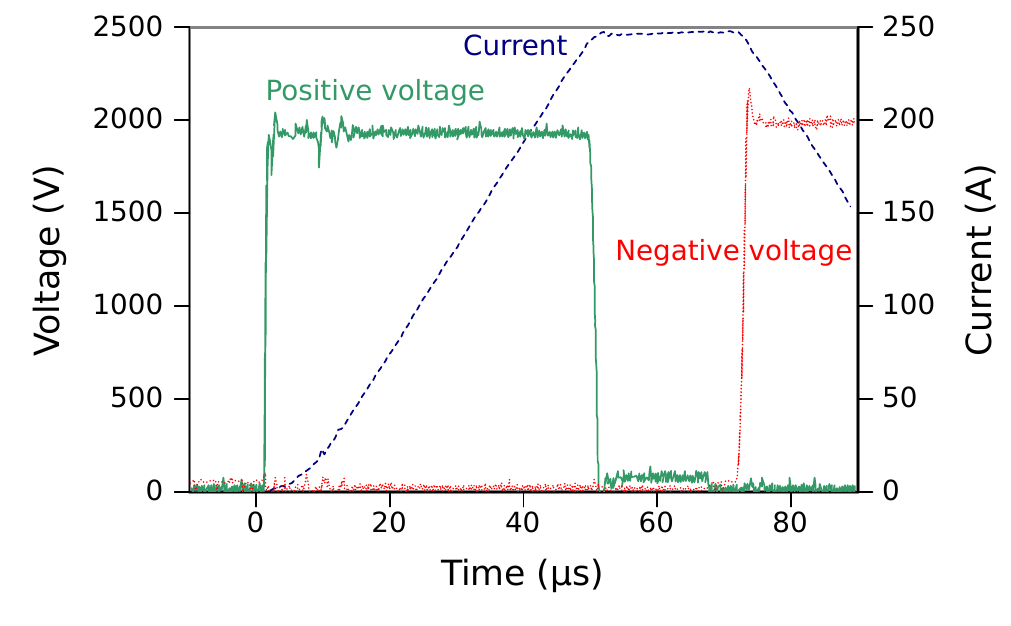} 
	\caption{Measured power-supply waveforms for 250~A demanded flat-top current, with 30~m of cable: Positive-side output voltage (green solid trace); negative-side output voltage (red dotted trace); output current (blue dashed trace).}
	\label{fig:normaloperation}
\end{figure}

Figure~\ref{fig:normaloperation} shows measured waveforms for 250\,A demanded flat-top current, with a 30\,m long cable between the power supply and magnet, for a flat-top duration of just over 20\,$\mu$s. The magnet current rise-time, 2\% to 98\%, is 46\,$\mu$s; however, the initial rate of rise of current is relatively low and becomes almost constant ($\sim 5.7$\,A/$\mu$s) after an elapsed time of 15\,$\mu$s. The positive-side output voltage (measured at node $P_{\rm{op}}$, with respect to ground, see Fig.~\ref{fig:powersupply}), to achieve this rate of rise of current is approximately 1930\,V, which corresponds to an inductance of approximately 338\,$\mu$H: it should be noted that the voltage $V_{\rm{HV}}$ across $C_{\rm{HV}}$ will be slightly higher than this due to resistance and inductance in the current path between $C_{\rm{HV}}$ and $P_{\rm{op}}$. During the rising edge of the current, the reduction in voltage across capacitor $C_{\rm{HV}}$ is approximately 30\,V (1.5\% of the initial voltage). Once the current reaches the required value, SW1 (Fig.~\ref{fig:powersupply}) is turned off, and SW3 is turned on and off, as necessary, to maintain the flat-top current. To generate the falling edge SW3 and SW2 are turned off; turning SW2 off forces the magnet current to freewheel through D2, and hence the negative-side output voltage (measured at node $N_{\rm{op}}$, with respect to ground, see Fig.~\ref{fig:powersupply}) rises to approximately 1980\,V: it should be noted that the voltage $V_{\rm{HV}}$ across $C_{\rm{HV}}$ will be slightly lower than this due to resistance and inductance in the current path between $C_{\rm{HV}}$ and $N_{\rm{op}}$. It should also be noted that the calibration of the power supply voltage dividers, at  $P_{\rm{op}}$ and $N_{\rm{op}}$ is $\pm 1$\%.

\begin{figure}[h]
	\includegraphics[width=\linewidth]{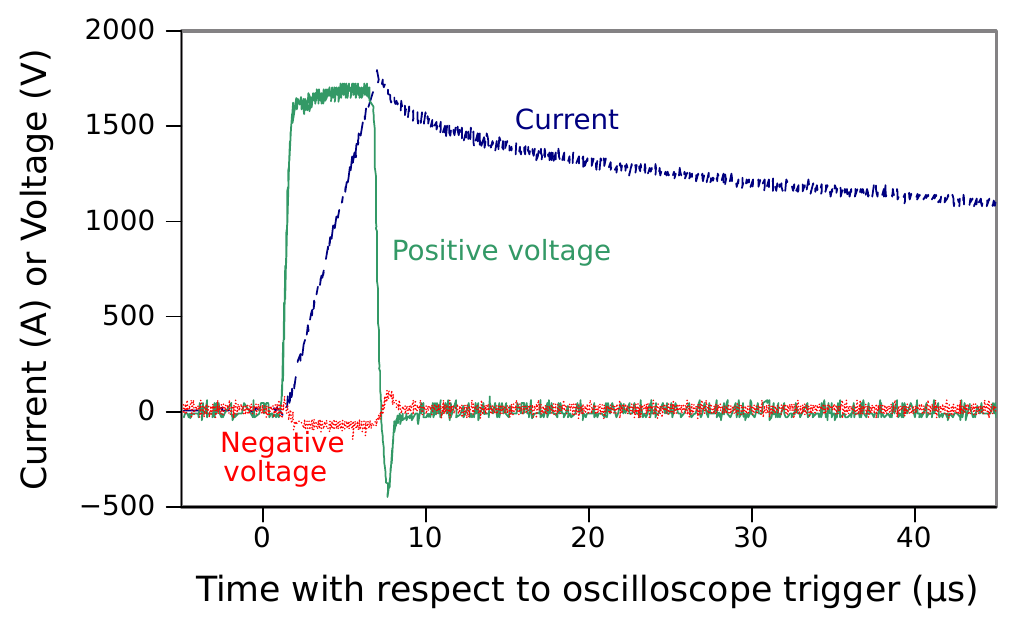} 
	\caption{Measured power supply waveforms for 250~A demanded flat-top current, with 30~m of cable and the magnet replaced with a short-circuit: Positive-side output voltage (green solid trace); negative-side output voltage (red dotted trace); output current (blue dashed trace).}
	\label{fig:shortcircuit}
\end{figure}

Several tests were carried out during the FATs where the magnet was replaced by a short-circuit.  These tests were to ensure that, in the case of a magnet fault, the power supply would safely ride-through the fault. The demanded current flat-top was gradually increased, proportionally increasing the high-voltage output of the power supply. Figure~\ref{fig:shortcircuit} shows the modulator output current for a demanded flat-top current of 250~A. The peak fault current is 1800~A: the delay for IGBT SW1 to turn off, after the 300~A over-current interlock is reached, is 4.5~$\mu$s. During the 4.5~$\mu$s, the average positive-side output voltage is approximately 1670~V, and the current increases by 1500~A: thus the inductance on the output of the modulator is $\sim$5.0~$\mu$H. Assuming that this inductance is predominantly due to the 30~m long cable, this corresponds to 167~nH/m for the two phase conductors.

Thermal cycling tests were also performed: the magnet was powered to 315~A dc for 10~minutes, then current was turned off for 10~minutes. This thermal cycle was repeated ten times. The voltage drop across the magnet, with 315~A dc, was measured to be 3.8~V. This corresponds to a maximum instantaneous total power dissipation of $\sim$1200~W and a dc resistance of 12.1~m$\Omega$ for the two saddle coils connected electrically in series. With a water flow rate of 7.6~l/minute per saddle coil the temperature rise across 6 turns, measured at the end of the 10~minutes with 315~A dc, was 1.4~$^\circ$C: the expected temperature rise was 1.1~$^\circ$C.

A 12~hour thermal test was carried out on the kicker magnet and power supply with 250~A, at 354~Hz and 50\% duty cycle: this corresponds to an rms current of $\sim$177~A and a predicted total power loss for the two saddle coils of almost 1200~W. The water flow was approximately 5.5~l/minute per saddle coil; the measured change in water temperature across a coil was 1.8$^\circ$C. The expected increase in water temperature, for 600~W dissipation per saddle coil, is 1.5$^\circ$C, indicating that the transient power dissipation is slightly higher than predicted. A thermal camera was used to measure temperatures in the power supply: the hottest components were an HV bleed resistor (45$^\circ$C), in parallel with ${C_{\rm{HV}}}$, and an LV bleed resistor (60$^\circ$C), in parallel with ${C_{\rm{LV}}}$. 


\subsection{Tests at TRIUMF without Beam\label{sec:prebeam}}
\begin{figure}[h]
\includegraphics[width=\linewidth]{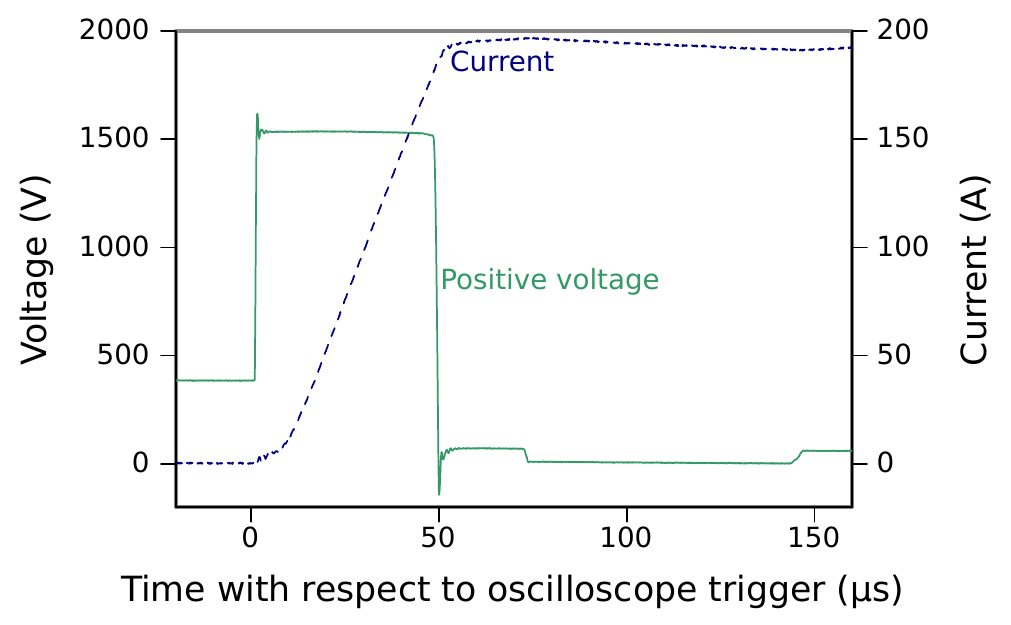} 
\caption{Measured rising edge of current pulse to an operating current of 193~A, subsequent to installation at TRIUMF. Dashed blue trace: current. Solid green trace: positive-side output voltage.}
\label{fig:risetime}
\end{figure}

Following installation of the power supply, magnet, and cabling at TRIUMF, further tests were carried out to ensure proper operation of controls, interlocks, and the system. The length of cable installed to connect power supply and magnet is $\sim$17~m. The blue dashed trace in Fig.~\ref{fig:risetime} is the power supply output current pulse at the nominal current of 193~A.  The measured rise-time, between 2\% and 98\%, is 47.2~$\mu$s---comfortably within the 50~$\mu$s specification. The overall delay, from the trigger pulse to 98\% of the current flat-top, is 52.0~$\mu$s. The value of this delay is required for proper synchronization with the beam pulse. The green solid trace in Fig.~\ref{fig:risetime} is the positive-side output voltage, which has a flat-top value of approximately 1540~V (${V_{\rm{HV}}}$) during the current rise-time.

\begin{figure}
\includegraphics[width=\linewidth]{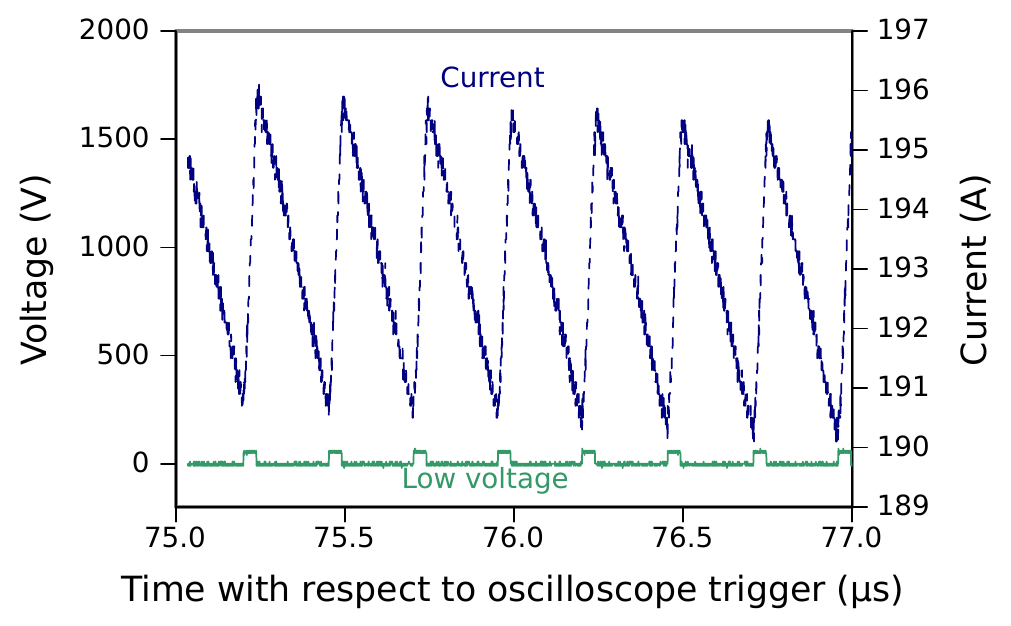} 
\caption{Measured flat top of current pulse at an operating current of 193~A. Blue dashed trace: current (ac-coupled measurement with an offset of 193~A added to the average value). Green solid trace: output voltage of low-voltage power supply stage to maintain flat-top current. }
\label{fig:flat-top}
\end{figure}

Figure~\ref{fig:flat-top} shows a zoom of the flat-top of the power supply output current pulse at the nominal current of 193~A (blue dashed trace). The ``saw-tooth" nature of the flat-top is created by the operation of the low-voltage stage of the power supply  (Fig.~\ref{fig:powersupply}).  The rising edge of the saw-tooth is created by switching on IGBT SW3: the green solid trace corresponds to $V_{\rm LV}$, and is ``high" when SW3 is on. Once the measured value of the magnet current is greater than a certain percentage above the demanded current SW3 is turned off, and the magnet current freewheels through diodes D3 and D1, and on-state 
IGBT SW2; the decay is due to circuit losses. Once the measured value of the magnet current is less than a certain percentage below the demanded current, SW3 is turned on again. The peak-to-peak amplitude of the saw-tooth current is 6~A, corresponding to $\pm 1.6$\% of the magnitude of the flat-top current, which is within the specified $\pm 2$\% (Table~\ref{tab:specifcations}).

\begin{figure}[h]
\includegraphics[width=\linewidth]{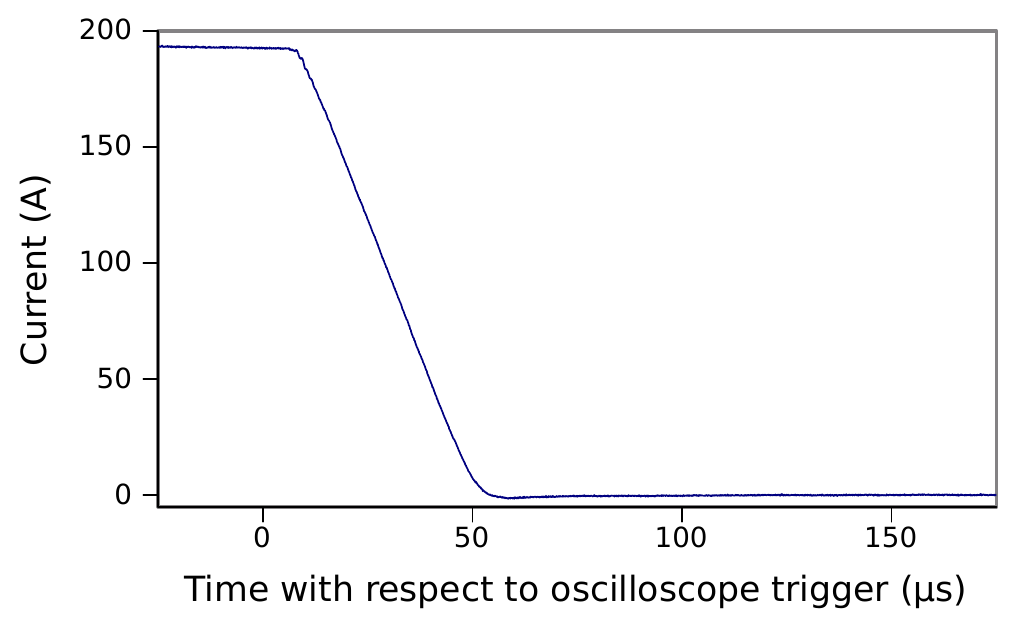} 
\caption{Measured falling edge of current pulse from 193~A flat-top.}
\label{fig:falltime}
\end{figure}

The trace in Fig.~\ref{fig:falltime} is the falling edge, from a flat-top of 193~A, of the power supply output current. The measured fall-time, between 98\% and 2\%, is 44.0~$\mu$s - comfortably within the 50~$\mu$s specification. The measured peak of the current undershoot is 1\% of the magnitude of the flat-top current, which is within the specified $\pm 2$\% for the post-pulse ripple (Table~\ref{tab:specifcations}).

\begin{figure}
  \includegraphics[width=\linewidth]{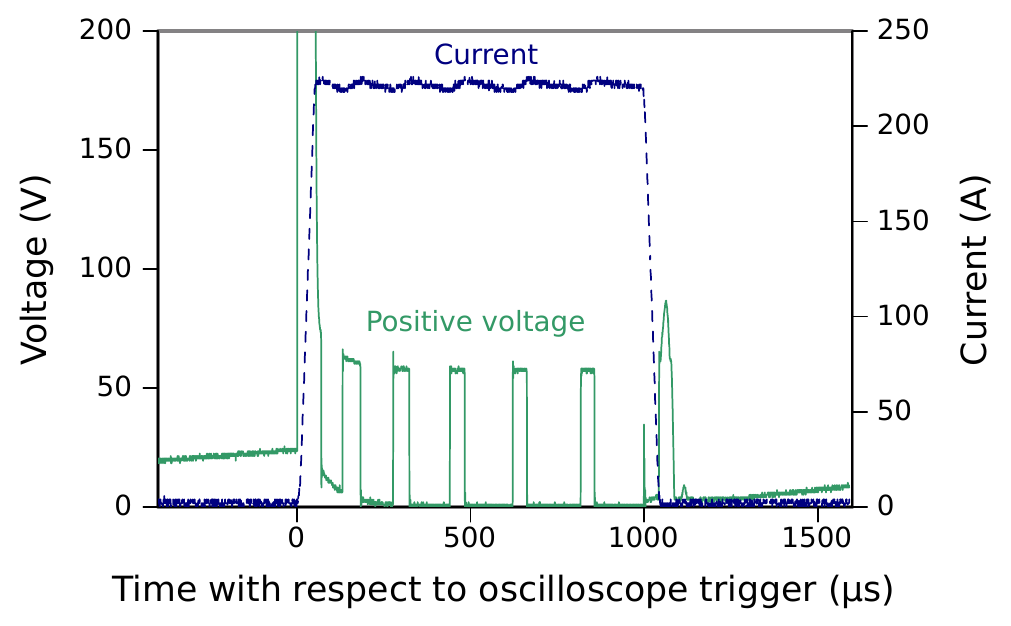} 
\caption{Measurement with 220~A current, pulsed at 350~Hz with a 1~ms-long flat-top. Dashed blue trace: current. Solid green trace: positive-side output voltage, zoomed in to show the low voltage which maintains the flat-top of the current pulse (described in the text).}
\label{fig:220A350Hz1ms}
\end{figure}

Tests and measurements were also carried out at 220~A output current at dc and pulsed at 350~Hz with a 1~ms-duration flat-top. Figure~\ref{fig:220A350Hz1ms} shows a measured current pulse (blue dashed) for the latter operating condition. The green solid trace in Fig.~\ref{fig:220A350Hz1ms} is the positive-side output voltage, which is off-scale during the rising edge of the current pulse. During the flat-top of the current pulse the low-voltage rectifier delivers output pulses of $\sim$60~V to maintain the magnitude of the flat-top of the current pulse to within the specified $\pm 2$\% (Table~\ref{tab:specifcations}). 

The magnet was installed into its position in beamline 1V as shown previously in Fig.~\ref{fig:1VUAschematic} with its shielding box on a custom stand. As mentioned earlier, a non-conducting vacuum tube must be installed in the aperture of the magnet.  A typical choice for the tube (described in Section~\ref{sec:kickermagnet}) is an alumina tube with a thin metallic coating to bleed off static charge deposited by beam halo.

The halo losses along this section were found to be relatively small ($\sim$1.5~pA/m).  This was determined by measuring the $^{56}$Co activity presumed to derive from $^{56}$Fe(p,n)$^{56}$Co in a new steel beam pipe that was inserted into the kicker section for the 2015 TRIUMF beam schedule, before the kicker magnet was installed.

Based in part on this measurement, an uncoated borosilicate glass tube was used for the first year of operation (2017).  After a year of use, this tube developed a leak in the O-ring seal in the transition to the metal beam pipe. Upon de-installation, the glass was found to have turned brown in color. This was expected from radiation effects in the glass, and generally does not affect the structural integrity of the glass.

The borosilicate glass tube was replaced with a fiberglass (FR-4) tube in early 2018. The fiberglass is expected to be more robust against possible stresses and more resistant to shatter.  In order to avoid O-ring failures caused by proton beam excursions, aluminum (6061-T6) flanges were epoxied onto the tube so that metal seals (1100 series aluminum) could be used. The fiberglass tube is also uncoated and thus far we have seen no evidence of any effects of the possible build up of static charge.

\section{Beam Delivery and Results}
\subsection{dc operation}
\label{sec:DC}

Low-intensity dc operation (kicker magnet supplied with a constant current) allows us to tune the beam by inserting HARP-style beam profile monitors and to make the necessary corrections to the currents of various other magnets.
The flat-top current of the kicker is also adjusted during this process to correct the vertical position of the beam at the septum.

The HARP monitors are designed for an average beam current below 50~nA. Due to their long integration time the time structure of this current is not important for HARP function. For example, kicking one pulse out of 2400 from a 120~$\mu$A beam in beamline 1V would deliver an adequate average current.
But the high instantaneous beam spill caused by the interaction with the HARP is sufficient to trigger beam spill monitors in beamline 1U.  So a lower instantaneous current is required.

At TRIUMF, the smaller currents in beamline 1V can be achieved by scraping the proton beam inside the cyclotron with the extraction foil, or by reducing the current injected into the cyclotron.
The former takes longer to tune but has the advantage that beam may be delivered to the other extraction ports without interruption.

\subsection{Kicker Controls and Diagnostics}
\label{sec:KSM}

As alluded to in relation to Fig.~\ref{fig:beam-sharing}, the H$^{-}$ injection system of the cyclotron utilizes a
pulser with a variable duty cycle, with a typical setting giving a notch duration of 50--100~$\mu$s. In our application, this notch is used to switch beam pulses between beamlines 1U and 1A by switching on or off the UCN kicker magnet current, respectively.  It is important that the UCN kicker control system ensures that the kicker magnet ramping is correctly timed within the notch. 

\begin{figure}[b]
\includegraphics[width=\linewidth]{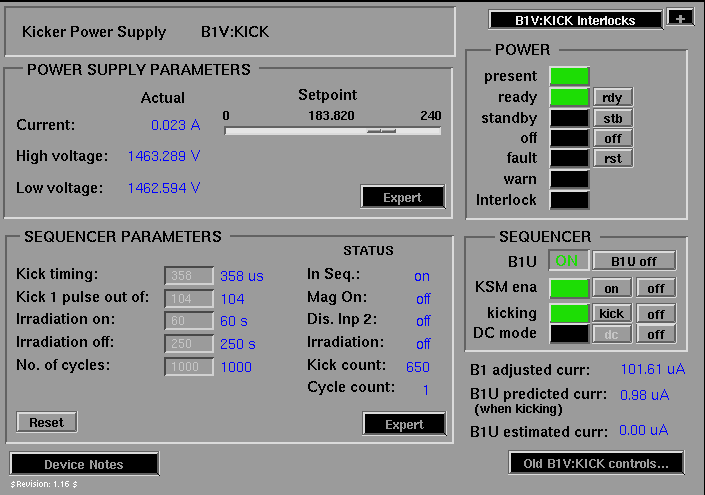}
\caption{EPICS control screen for kicker magnet showing controls for the power supply (top right), controls for the sequencer module (bottom right), flat-top magnet current (top left), and timing parameters (bottom left).}
\label{fig:ksm_epics_control}
\end{figure}

\begin{figure}
    \centering
    \includegraphics[width=\linewidth]{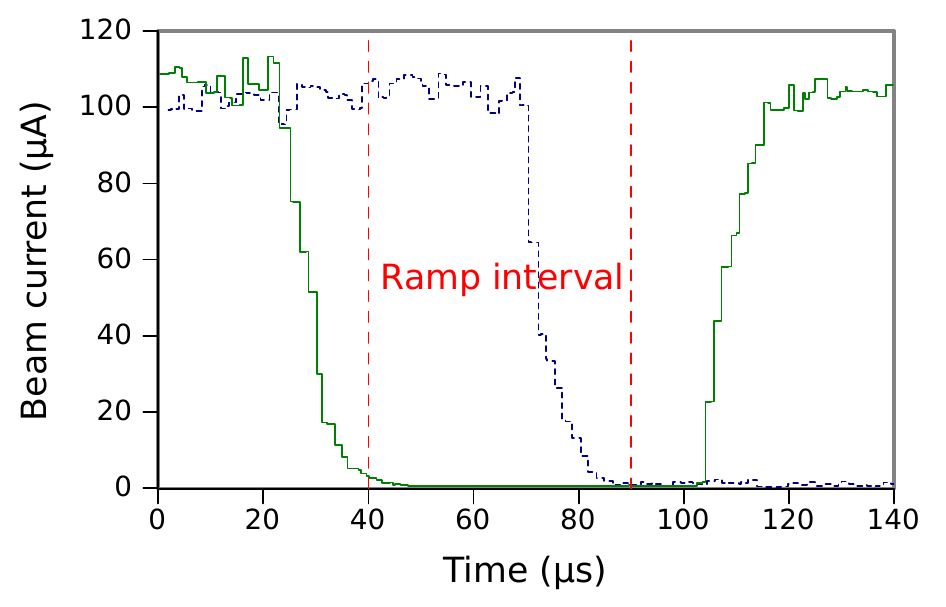}
    \caption{Beam current in beamline 1V as measured with the 1VM4.7 capacitive probe. If the notch is correctly aligned (green solid line) with the time when the kicker magnet is ramping (vertical red lines) no beam is spilled on the septum magnet. A misaligned notch (blue dashed line) will cause significant beam spill.}
    \label{fig:ktm_plots}
\end{figure}

The first element of the kicker controls is the Kicker Sequencer Module (KSM), which is a TRIUMF-built VME module.  The KSM receives a signal from the H$^{-}$ pulser indicating when the ions are injected into the cyclotron. The KSM must add a delay to this signal ($\sim$320~$\mu$s) to account for the travel time of the protons through the cyclotron. The delay must be variable, because the travel time varies depending on the cyclotron operating mode.  At this delayed injection time the KSM sends a signal to the UCN kicker power supply to either ramp up or down the kicker magnet current.  The KSM also sets the fraction of beam pulses delivered to beamline 1U and hence its average beam current.  The KSM and the kicker power supply itself are controlled using EPICS~\cite{bib:epics}, and a picture of the control screen is shown in Fig.~\ref{fig:ksm_epics_control}.

In the final configuration one beam pulse out of each three will be directed to the UCN beamline, achieving 40~$\mu$A of proton beam average current.  We also plan to implement the ability to kick more general patterns of pulses, to adjust the average beamline 1U current more precisely.  For instance, we will eventually have the ability to deliver 3 beam pulses out of 10 to beamline 1U, which is not achievable with our current software.
  
The second element of the kicker controls is a fast diagnostic element to measure the beam notch.  The diagnostic element is a capacitive probe in beamline 1V upstream of the septum magnet; it is designated as 1VM4.7.  The signal from the 1VM4.7 probe is shaped, digitized with a giga-sample per second ADC, and then analyzed to calculate the beam power as a function of time.  Examples of the 1VM4.7 waveforms are shown in Fig.~\ref{fig:ktm_plots}.  These waveforms are used to adjust the KSM delay time so that the kicker magnet ramping time is aligned well within the beam notch.

The 1VM4.7 measurements can also be used to measure how clean (absence of beam current) the beam notch is.  Depending on the cyclotron tune it is possible that the notch is correctly aligned in time, but that there is significant beam still in the notch.  The 1VM4.7 measurements allow the cyclotron operators to check for such conditions before kicking beam.

The procedure for setting up to kick beam to the UCN beamline involves adjusting the KSM delay time until the magnet current ramp-up timing is well aligned with the cyclotron notch.  We have found in practice that there is relatively little variation in the travel time through the cyclotron for the normal cyclotron operating modes, at least compared to the duration of the beam notch.  So it is only rarely necessary to adjust the KSM delay time.

It should also be noted that the other diagnostic devices in the UCN beamline (such as the beam current, beam position and target protection monitors) need to have electronics that can handle the different beam duty cycles that the kicker can provide.  This is discussed further in Section~\ref{sec:experiences}.

\subsection{Results of Kicker Operation with Proton Beam}
\label{sec:mistiming}

\begin{figure}
\includegraphics[width=\linewidth]{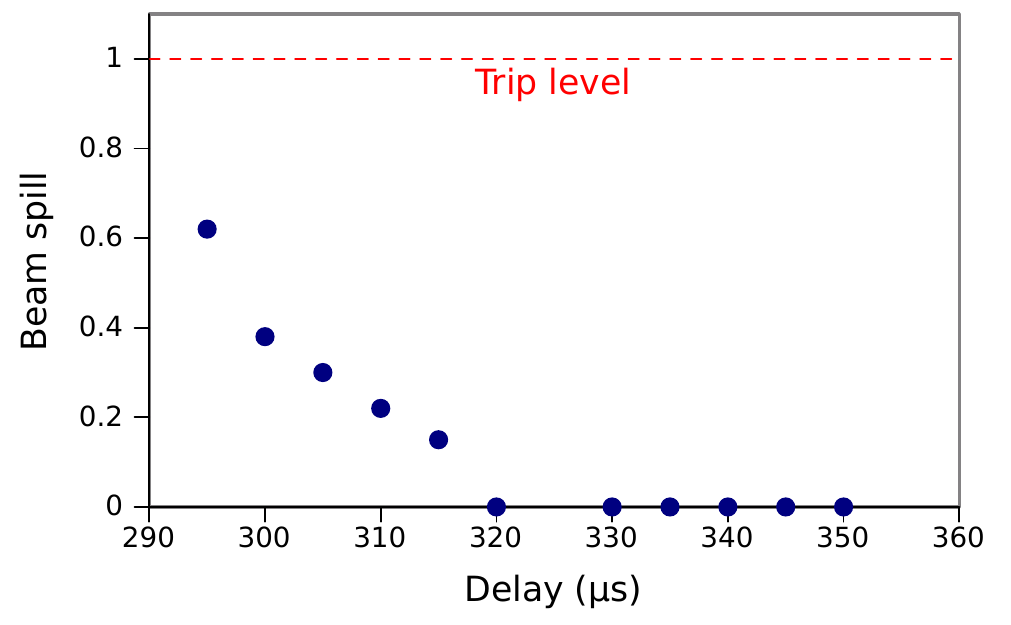}
\caption{Signal of beam spill monitor downstream of septum magnet, as a function of the kicker delay with respect to the cyclotron injection pulser.  The beam spill signal is expressed as a fraction of the value at which the beam spill monitor will trip the cyclotron.}
\label{fig:ktm_mistimg}
\end{figure}

In summer 2017 the kicker magnet system was operated in a kicking mode with beams simultaneously delivered to beamlines 1U and 1A for the first time.  One of the first experiments carried out was to investigate the effect of poorly aligning the magnet ramp with the beam notch.  Fig.~\ref{fig:ktm_plots} shows examples from correctly aligned and poorly aligned magnet ramp signals. Kicking one out of four 1~$\mu$A beam pulses in beamline 1V, we measured the signal from a beam spill monitor located downstream of the UCN septum magnet while changing the time delay.  The results of the test are shown in Fig.~\ref{fig:ktm_mistimg}, which confirms the expected behaviour; more beam is spilled as the magnet ramp signal becomes more misaligned with the beam notch.  The results show that the 1VM4.7 measurement can be used to accurately set the kicker ramp time and that poor alignment of the magnet ramp and the beam notch are correctly detected by the beam spill monitors (note that there is a fixed time delay for the digitizer measurement in Fig.~\ref{fig:ktm_plots}, which is why the timescale is different from  Fig.~\ref{fig:ktm_mistimg}).

Once the magnet ramp was properly aligned with the notch no beam spill was measurable, confirming that the magnet can ramp within the notch and the measured maximum ripple amplitude of 1.6\% is acceptable. However, to compensate slight variations in the arrival time of the notch we had to increase the minimum width of the notch to about 60\,$\mu$s. This slightly limits the duty cycle of the injection pulser to below 93\% and the cyclotron's ability to adjust the beam power.  Generally this poses no operational concern for the other cyclotron users.

\subsection{Implementation and Experiences}
\label{sec:experiences}

As of this writing, beamline 1U has been operated in kicking mode up to $10\,\mu$A.
Kick fractions between 1/10,000 and 1/4, corresponding to kicker frequencies of 0.1~Hz to 282~Hz, have been realized without any magnet or power supply issues.

\begin{figure}
\includegraphics[width=\linewidth]{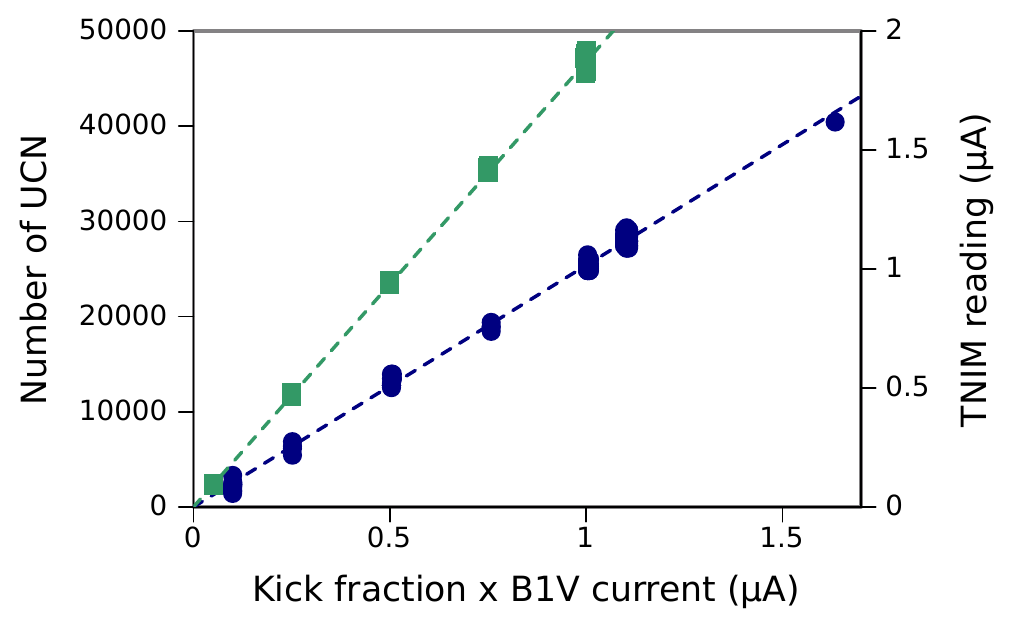}
\caption{The current measured by the TNIM (circles) and the number of UCN extracted from the UCN source (boxes) increase linearly with the beam current calculated from the kicker parameters and the current in beamline 1V.}
\label{fig:TNIMreading}
\end{figure}

A prototype UCN source, originally developed in Japan~\cite{PhysRevLett.108.134801}, was installed above the spallation target in 2017. 
Two month-long experimental campaigns were performed with the UCN source in Fall 2017 and 2018. 
The source was previously operated with a nominal beam current of 1~$\mu$A. Hence, we typically kicked every hundredth pulse from a 100~$\mu$A beam in beamline 1V, into 1U, to achieve this current. The relatively sparse pulsing of the beam posed a challenge to the toroidal non-intercepting monitor (TNIM) which measures beam current.  The electronics for the TNIM relied on the 1~kHz structure of the beam to make a precise measurement of the current.
The TNIM required a new readout, and this was based on digital filters implemented on an FPGA~\cite{bib:rawnsley}.
With this new readout, we were able to calibrate the TNIM and use it to confirm that the beam current in beamline 1U was proportional to the fraction of pulses kicked out of beamline 1V multiplied with the beam current in beamline 1V, as expected (Fig.~\ref{fig:TNIMreading}).

During the Fall 2017 campaign~\cite{UCNprod}, we also confirmed that the number of UCN extracted from the UCN source is proportional to the calculated beam current (Fig.~\ref{fig:TNIMreading}).  In these measurements, the target was normally irradiated for 60~s using the kicker magnet.  During the beam-on period, UCN were produced and stored in the source.  When the irradiation was complete, a valve was opened and UCN were transported to a detector.  The detected number of UCN was found to be proportional up to beam currents of 1~$\mu$A, as shown in Fig.~\ref{fig:TNIMreading}. Measurements up to 10~$\mu$A of average proton current are reported in Ref.~\cite{UCNprod}; the number of detected UCN falls below linear extrapolation, which is an indication that the cooling power of the prototype UCN source is insufficient.  The UCN source will be upgraded to make use of the full 40~$\mu$A design current provided by the kicker magnet.

During the beam periods mentioned above, beamline 1U was normally retuned after an interruption of beam for more than a few hours.
Since the tuning requires low-current injection into the cyclotron, this caused downtime for other beamlines on other extraction ports, disrupting beam delivery to other experiments at TRIUMF.
In 2018 the stability of the beamline was demonstrated to be sufficient to avoid retuning, even after interruptions of 12~hours or longer.
Procedures were then changed for the month-long Fall 2018 UCN experiment campaign.  In the new procedures, magnet power supplies for the beamline 1U optics elements were kept at current and kicking was restored on demand for the UCN experiments.
This form of operation was successful for the duration of the experiments, with no excessive beam spill being observed.

\section{Conclusion}
A 12-turn kicker magnet has been designed and operated at TRIUMF, providing proton beam to a new UCN source facility. The 12 turns are a compromise between requiring low current, to achieve the required deflection of the beam with low power loss in the coil, and the need to supply high voltage for relatively fast field rise and fall times. In addition, a power supply based on IGBTs has been designed and operated together with the kicker magnet. The design of the power supply and rating of components are conservative to ensure high reliability. The power supply consists of two stages, one high-voltage part to achieve the fast rise and fall times, and a low-voltage part to enable a selectable output pulse flat-top duration from 0~s to dc. Factory acceptance tests and tests at TRIUMF demonstrate that the required rise and fall times are achieved, as are the  flat-top ripple of $\leq\pm 2$\%, field homogeneity $\leq\pm 2$\%, and a continuous repetition rate selectable from dc to 400~Hz.

In 2017 and 2018 the kicker magnet system was operated reliably over two month-long periods while UCN production experiments were being carried out. The magnet delivered up to 10~$\mu$A of instantaneous current and a total 110~$\mu$A$\cdot$h of integrated current to the spallation target.  The sensed beam current in the TNIM was found to be linear with the fraction of beam pulses delivered to beamline 1U. The UCN production results were also found to be linear with the beam current, up to the heat load acceptable for good UCN source operation.  Tests where the timing of the ramping of the kicker magnet was purposely misaligned generated beam spills as expected. During normal operation no excessive beam spill was detected, confirming that the requirements of the kicker system were properly specified and met. The timing with the beam notch could be easily adjusted and monitored with a capacitive probe. In implementing the system, we found several challenges, particularly with respect to diagnostic tools which were designed for a more continuous current.  We expect that the kicker system will fully serve the needs of the future UCN source facility which, once upgraded, will require regular delivery of up to 40~$\mu$A time-averaged proton current.

\begin{acknowledgments}
The authors thank C.~Marshall, C.A.~Miller, J.~Pon, D.~Preddy, and C.A.~Remon for their important contributions to this work.
This work was undertaken, in part, thanks to funding from the Natural
Sciences and Engineering Research Council Canada, the Canada Research
Chairs program, and the Canada Foundation for Innovation.
\end{acknowledgments}

\bibliography{mybibfile}

\begin{thebibliography}{36}%
\makeatletter
\providecommand \@ifxundefined [1]{%
 \@ifx{#1\undefined}
}%
\providecommand \@ifnum [1]{%
 \ifnum #1\expandafter \@firstoftwo
 \else \expandafter \@secondoftwo
 \fi
}%
\providecommand \@ifx [1]{%
 \ifx #1\expandafter \@firstoftwo
 \else \expandafter \@secondoftwo
 \fi
}%
\providecommand \natexlab [1]{#1}%
\providecommand \enquote  [1]{``#1''}%
\providecommand \bibnamefont  [1]{#1}%
\providecommand \bibfnamefont [1]{#1}%
\providecommand \citenamefont [1]{#1}%
\providecommand \href@noop [0]{\@secondoftwo}%
\providecommand \href [0]{\begingroup \@sanitize@url \@href}%
\providecommand \@href[1]{\@@startlink{#1}\@@href}%
\providecommand \@@href[1]{\endgroup#1\@@endlink}%
\providecommand \@sanitize@url [0]{\catcode `\\12\catcode `\$12\catcode
  `\&12\catcode `\#12\catcode `\^12\catcode `\_12\catcode `\%12\relax}%
\providecommand \@@startlink[1]{}%
\providecommand \@@endlink[0]{}%
\providecommand \url  [0]{\begingroup\@sanitize@url \@url }%
\providecommand \@url [1]{\endgroup\@href {#1}{\urlprefix }}%
\providecommand \urlprefix  [0]{URL }%
\providecommand \Eprint [0]{\href }%
\providecommand \doibase [0]{http://dx.doi.org/}%
\providecommand \selectlanguage [0]{\@gobble}%
\providecommand \bibinfo  [0]{\@secondoftwo}%
\providecommand \bibfield  [0]{\@secondoftwo}%
\providecommand \translation [1]{[#1]}%
\providecommand \BibitemOpen [0]{}%
\providecommand \bibitemStop [0]{}%
\providecommand \bibitemNoStop [0]{.\EOS\space}%
\providecommand \EOS [0]{\spacefactor3000\relax}%
\providecommand \BibitemShut  [1]{\csname bibitem#1\endcsname}%
\let\auto@bib@innerbib\@empty
\bibitem [{\citenamefont {Golub}\ and\ \citenamefont
  {Pendlebury}(1975)}]{GOLUB1975133}%
  \BibitemOpen
  \bibfield  {author} {\bibinfo {author} {\bibfnamefont {R.}~\bibnamefont
  {Golub}}\ and\ \bibinfo {author} {\bibfnamefont {J.}~\bibnamefont
  {Pendlebury}},\ }\href {\doibase 10.1016/0375-9601(75)90500-9} {\bibfield
  {journal} {\bibinfo  {journal} {Phys. Lett. A}\ }\textbf {\bibinfo {volume}
  {53}},\ \bibinfo {pages} {133 } (\bibinfo {year} {1975})}\BibitemShut
  {NoStop}%
\bibitem [{\citenamefont {Schmidt-Wellenburg}\ \emph
  {et~al.}(2015)\citenamefont {Schmidt-Wellenburg}, \citenamefont {Bossy},
  \citenamefont {Farhi}, \citenamefont {Fertl}, \citenamefont {Leung},
  \citenamefont {Rahli}, \citenamefont {Soldner},\ and\ \citenamefont
  {Zimmer}}]{PhysRevC.92.024004}%
  \BibitemOpen
  \bibfield  {author} {\bibinfo {author} {\bibfnamefont {P.}~\bibnamefont
  {Schmidt-Wellenburg}}, \bibinfo {author} {\bibfnamefont {J.}~\bibnamefont
  {Bossy}}, \bibinfo {author} {\bibfnamefont {E.}~\bibnamefont {Farhi}},
  \bibinfo {author} {\bibfnamefont {M.}~\bibnamefont {Fertl}}, \bibinfo
  {author} {\bibfnamefont {K.~K.~H.}\ \bibnamefont {Leung}}, \bibinfo {author}
  {\bibfnamefont {A.}~\bibnamefont {Rahli}}, \bibinfo {author} {\bibfnamefont
  {T.}~\bibnamefont {Soldner}}, \ and\ \bibinfo {author} {\bibfnamefont
  {O.}~\bibnamefont {Zimmer}},\ }\href {\doibase 10.1103/PhysRevC.92.024004}
  {\bibfield  {journal} {\bibinfo  {journal} {Phys. Rev. C}\ }\textbf {\bibinfo
  {volume} {92}},\ \bibinfo {pages} {024004} (\bibinfo {year}
  {2015})}\BibitemShut {NoStop}%
\bibitem [{\citenamefont {Korobkina}\ \emph {et~al.}(2002)\citenamefont
  {Korobkina} \emph {et~al.}}]{KOROBKINA2002462}%
  \BibitemOpen
  \bibfield  {author} {\bibinfo {author} {\bibfnamefont {E.}~\bibnamefont
  {Korobkina}} \emph {et~al.},\ }\href {\doibase 10.1016/S0375-9601(02)01052-6}
  {\bibfield  {journal} {\bibinfo  {journal} {Phys. Lett. A}\ }\textbf
  {\bibinfo {volume} {301}},\ \bibinfo {pages} {462 } (\bibinfo {year}
  {2002})}\BibitemShut {NoStop}%
\bibitem [{\citenamefont {Dilling}\ \emph {et~al.}(2014)\citenamefont
  {Dilling}, \citenamefont {Kr\"{u}cken},\ and\ \citenamefont
  {Merminga}}]{ISACandARIEL}%
  \BibitemOpen
  \bibinfo {editor} {\bibfnamefont {J.}~\bibnamefont {Dilling}}, \bibinfo
  {editor} {\bibfnamefont {R.}~\bibnamefont {Kr\"{u}cken}}, \ and\ \bibinfo
  {editor} {\bibfnamefont {L.}~\bibnamefont {Merminga}},\ eds.,\ \href
  {\doibase 10.1007/978-94-007-7963-1} {\emph {\bibinfo {title} {{ISAC} and
  {ARIEL}: The {TRIUMF} Radioactive Beam Facilities and the Scientific
  Program}}}\ (\bibinfo  {publisher} {Springer Netherlands},\ \bibinfo {year}
  {2014})\BibitemShut {NoStop}%
\bibitem [{\citenamefont {Barnes}(2018)}]{Barnes:CAS2017}%
  \BibitemOpen
  \bibfield  {author} {\bibinfo {author} {\bibfnamefont {M.~J.}\ \bibnamefont
  {Barnes}},\ }in\ \href {\doibase
  http://dx.doi.org/10.23730/CYRSP-2018-005.229} {\emph {\bibinfo {booktitle}
  {Proceedings of the CAS-CERN Accelerator School: Beam Injection, Extraction
  and Transfer}}}\ (\bibinfo {address}
  {http://dx.doi.org/10.23730/CYRSP-2018-005.229},\ \bibinfo {year} {2018})\
  pp.\ \bibinfo {pages} {229--283}\BibitemShut {NoStop}%
\bibitem [{\citenamefont {Ahmed}\ \emph
  {et~al.}(2019{\natexlab{a}})\citenamefont {Ahmed} \emph
  {et~al.}}]{AHMED2019101}%
  \BibitemOpen
  \bibfield  {author} {\bibinfo {author} {\bibfnamefont {S.}~\bibnamefont
  {Ahmed}} \emph {et~al.},\ }\href {\doibase
  https://doi.org/10.1016/j.nima.2019.01.074} {\bibfield  {journal} {\bibinfo
  {journal} {Nuclear Instruments and Methods in Physics Research Section A:
  Accelerators, Spectrometers, Detectors and Associated Equipment}\ }\textbf
  {\bibinfo {volume} {927}},\ \bibinfo {pages} {101 } (\bibinfo {year}
  {2019}{\natexlab{a}})}\BibitemShut {NoStop}%
\bibitem [{\citenamefont {Wang}(2019)}]{bib:rwang}%
  \BibitemOpen
  \bibfield  {author} {\bibinfo {author} {\bibfnamefont {R.}~\bibnamefont
  {Wang}},\ }\href@noop {} {}\bibinfo {howpublished} {{Hons.~thesis, The
  University of British Columbia}} (\bibinfo {year} {2019})\BibitemShut
  {NoStop}%
\bibitem [{\citenamefont {Ahmed}\ \emph
  {et~al.}(2019{\natexlab{b}})\citenamefont {Ahmed} \emph {et~al.}}]{UCNprod}%
  \BibitemOpen
  \bibfield  {author} {\bibinfo {author} {\bibfnamefont {S.}~\bibnamefont
  {Ahmed}} \emph {et~al.} (\bibinfo {collaboration} {TUCAN Collaboration}),\
  }\href {\doibase 10.1103/PhysRevC.99.025503} {\bibfield  {journal} {\bibinfo
  {journal} {Phys. Rev. C}\ }\textbf {\bibinfo {volume} {99}},\ \bibinfo
  {pages} {025503} (\bibinfo {year} {2019}{\natexlab{b}})}\BibitemShut
  {NoStop}%
\bibitem [{\citenamefont {Anicic}\ \emph {et~al.}(2005)\citenamefont {Anicic}
  \emph {et~al.}}]{PSI:Anicic2005}%
  \BibitemOpen
  \bibfield  {author} {\bibinfo {author} {\bibfnamefont {D.}~\bibnamefont
  {Anicic}} \emph {et~al.},\ }\href {\doibase doi:10.1016/j.nima.2004.12.032}
  {\bibfield  {journal} {\bibinfo  {journal} {Nucl. Instrum. Methods Phys.
  Res., Sect. A}\ }\textbf {\bibinfo {volume} {541}},\ \bibinfo {pages} {598
  –609} (\bibinfo {year} {2005})}\BibitemShut {NoStop}%
\bibitem [{\citenamefont {Pullia}(2006)}]{CNAO:Pullia2005}%
  \BibitemOpen
  \bibfield  {author} {\bibinfo {author} {\bibfnamefont {M.}~\bibnamefont
  {Pullia}},\ }\href {\doibase 10.1109/TASC.2005.869681} {\bibfield  {journal}
  {\bibinfo  {journal} {IEEE Transactions on Applied Superconductivity}\
  }\textbf {\bibinfo {volume} {16}},\ \bibinfo {pages} {1708} (\bibinfo {year}
  {2006})}\BibitemShut {NoStop}%
\bibitem [{\citenamefont {Benedikt}(2005)}]{medaustron:Benedikt2005}%
  \BibitemOpen
  \bibfield  {author} {\bibinfo {author} {\bibfnamefont {M.}~\bibnamefont
  {Benedikt}},\ }\href {\doibase 10.1016/j.nima.2004.09.038} {\bibfield
  {journal} {\bibinfo  {journal} {Nucl. Instrum. Methods Phys. Res., Sect. A}\
  }\textbf {\bibinfo {volume} {539}},\ \bibinfo {pages} {25 } (\bibinfo {year}
  {2005})}\BibitemShut {NoStop}%
\bibitem [{\citenamefont {Benedikt}\ \emph {et~al.}(2010)\citenamefont
  {Benedikt}, \citenamefont {Gutleber}, \citenamefont {Palm}, \citenamefont
  {Pirkl}, \citenamefont {Dorda},\ and\ \citenamefont
  {Fabich}}]{medaustron:Benedikt2010}%
  \BibitemOpen
  \bibfield  {author} {\bibinfo {author} {\bibfnamefont {M.}~\bibnamefont
  {Benedikt}}, \bibinfo {author} {\bibfnamefont {J.}~\bibnamefont {Gutleber}},
  \bibinfo {author} {\bibfnamefont {M.}~\bibnamefont {Palm}}, \bibinfo {author}
  {\bibfnamefont {W.}~\bibnamefont {Pirkl}}, \bibinfo {author} {\bibfnamefont
  {U.}~\bibnamefont {Dorda}}, \ and\ \bibinfo {author} {\bibfnamefont
  {A.}~\bibnamefont {Fabich}},\ }in\ \href@noop {} {\emph {\bibinfo {booktitle}
  {IPAC'10}}}\ (\bibinfo {address} {Kyoto, Japan},\ \bibinfo {year} {2010})\
  pp.\ \bibinfo {pages} {109--111}\BibitemShut {NoStop}%
\bibitem [{\citenamefont {Sermeus}\ \emph {et~al.}(2004)\citenamefont
  {Sermeus}, \citenamefont {Borburgh}, \citenamefont {Fowler}, \citenamefont
  {Hourican}, \citenamefont {Metzmacher},\ and\ \citenamefont
  {Crescenti}}]{Sermeus2004}%
  \BibitemOpen
  \bibfield  {author} {\bibinfo {author} {\bibfnamefont {L.}~\bibnamefont
  {Sermeus}}, \bibinfo {author} {\bibfnamefont {J.}~\bibnamefont {Borburgh}},
  \bibinfo {author} {\bibfnamefont {A.}~\bibnamefont {Fowler}}, \bibinfo
  {author} {\bibfnamefont {M.}~\bibnamefont {Hourican}}, \bibinfo {author}
  {\bibfnamefont {K.~D.}\ \bibnamefont {Metzmacher}}, \ and\ \bibinfo {author}
  {\bibfnamefont {M.}~\bibnamefont {Crescenti}},\ }in\ \href@noop {} {\emph
  {\bibinfo {booktitle} {EPAC'04}}}\ (\bibinfo {address} {Lucerne,
  Switzerland},\ \bibinfo {year} {2004})\ pp.\ \bibinfo {pages}
  {1639--1641}\BibitemShut {NoStop}%
\bibitem [{\citenamefont {Dallago}\ \emph {et~al.}(2006)\citenamefont {Dallago}
  \emph {et~al.}}]{Dallago2006}%
  \BibitemOpen
  \bibfield  {author} {\bibinfo {author} {\bibfnamefont {E.}~\bibnamefont
  {Dallago}} \emph {et~al.},\ }in\ \href@noop {} {\emph {\bibinfo {booktitle}
  {2006 37th IEEE Power Electronics Specialists Conference}}}\ (\bibinfo
  {address} {Jeju, South Korea},\ \bibinfo {year} {2006})\BibitemShut {NoStop}%
\bibitem [{\citenamefont {Borburgh}\ \emph {et~al.}(2010)\citenamefont
  {Borburgh} \emph {et~al.}}]{Borburgh2010}%
  \BibitemOpen
  \bibfield  {author} {\bibinfo {author} {\bibfnamefont {J.}~\bibnamefont
  {Borburgh}} \emph {et~al.},\ }in\ \href@noop {} {\emph {\bibinfo {booktitle}
  {IPAC'10}}}\ (\bibinfo {address} {Kyoto, Japan},\ \bibinfo {year} {2010})\
  pp.\ \bibinfo {pages} {3954--3956}\BibitemShut {NoStop}%
\bibitem [{\citenamefont {Stadlbauer}\ \emph {et~al.}(2015)\citenamefont
  {Stadlbauer} \emph {et~al.}}]{Stadlbauer2015}%
  \BibitemOpen
  \bibfield  {author} {\bibinfo {author} {\bibfnamefont {T.}~\bibnamefont
  {Stadlbauer}} \emph {et~al.},\ }in\ \href@noop {} {\emph {\bibinfo
  {booktitle} {IPAC'15}}}\ (\bibinfo {address} {Richmond, USA},\ \bibinfo
  {year} {2015})\ pp.\ \bibinfo {pages} {2741--2743}\BibitemShut {NoStop}%
\bibitem [{\citenamefont {Bryant}\ \emph {et~al.}(2000)\citenamefont {Bryant}
  \emph {et~al.}}]{Bryant2000}%
  \BibitemOpen
  \bibfield  {author} {\bibinfo {author} {\bibfnamefont {P.~J.}\ \bibnamefont
  {Bryant}} \emph {et~al.} (\bibinfo {collaboration} {Accelerator Complex Study
  Group})\ }(\bibinfo {year} {2000})\BibitemShut {NoStop}%
\bibitem [{\citenamefont {Kramer}\ \emph {et~al.}(2011)\citenamefont {Kramer},
  \citenamefont {Stadlbauer}, \citenamefont {Barnes}, \citenamefont
  {Benedikt},\ and\ \citenamefont {Fowler}}]{Kramer2011}%
  \BibitemOpen
  \bibfield  {author} {\bibinfo {author} {\bibfnamefont {T.}~\bibnamefont
  {Kramer}}, \bibinfo {author} {\bibfnamefont {T.}~\bibnamefont {Stadlbauer}},
  \bibinfo {author} {\bibfnamefont {M.~J.}\ \bibnamefont {Barnes}}, \bibinfo
  {author} {\bibfnamefont {M.}~\bibnamefont {Benedikt}}, \ and\ \bibinfo
  {author} {\bibfnamefont {T.}~\bibnamefont {Fowler}},\ }in\ \href@noop {}
  {\emph {\bibinfo {booktitle} {IPAC'11}}}\ (\bibinfo {address} {San Sebastian,
  Spain},\ \bibinfo {year} {2011})\ pp.\ \bibinfo {pages}
  {3380--3390}\BibitemShut {NoStop}%
\bibitem [{\citenamefont {Atanasov}(2014)}]{Miro-thesis}%
  \BibitemOpen
  \bibfield  {author} {\bibinfo {author} {\bibfnamefont {M.}~\bibnamefont
  {Atanasov}},\ }\emph {\bibinfo {title} {{Design Optimization of the Fast
  Switched Chopper Dipole Magnet for the MedAustron Project}}},\ \href@noop {}
  {Master's thesis},\ \bibinfo  {school} {{Technical University of Sofia,
  Plovdiv Branch}} (\bibinfo {year} {2014})\BibitemShut {NoStop}%
\bibitem [{\citenamefont {Shoji}\ \emph {et~al.}(2010)\citenamefont {Shoji}
  \emph {et~al.}}]{Shoji}%
  \BibitemOpen
  \bibfield  {author} {\bibinfo {author} {\bibfnamefont {T.}~\bibnamefont
  {Shoji}} \emph {et~al.},\ }in\ \href {\doibase 10.1109/IPEC.2010.5543845}
  {\emph {\bibinfo {booktitle} {The 2010 International Power Electronics
  Conference}}}\ (\bibinfo {address} {Sapporo, Japan},\ \bibinfo {year}
  {2010})\ pp.\ \bibinfo {pages} {142--148}\BibitemShut {NoStop}%
\bibitem [{\citenamefont {{Dassault Systems, Opera Simulation Software,
  Kidlington, Oxfordshire, UK}}(2019)}]{cobham}%
  \BibitemOpen
  \bibfield  {author} {\bibinfo {author} {\bibnamefont {{Dassault Systems,
  Opera Simulation Software, Kidlington, Oxfordshire, UK}}},\ }\href@noop {}
  {}\bibinfo {howpublished} {\url{https://operafea.com}} (\bibinfo {year}
  {2019})\BibitemShut {NoStop}%
\bibitem [{\citenamefont {Hahn}(2011)}]{Hahn}%
  \BibitemOpen
  \bibfield  {author} {\bibinfo {author} {\bibfnamefont {M.}~\bibnamefont
  {Hahn}},\ }\href@noop {} {}\bibinfo {howpublished} {{Bachelor's thesis,
  Coburg University of Applied Sciences and Arts}} (\bibinfo {year}
  {2011})\BibitemShut {NoStop}%
\bibitem [{\citenamefont {{Luvata, FI-28330 Pori, Finland}}(2013)}]{Luvata}%
  \BibitemOpen
  \bibfield  {author} {\bibinfo {author} {\bibnamefont {{Luvata, FI-28330 Pori,
  Finland}}},\ }\href@noop {} {}\bibinfo {howpublished}
  {\url{http://www.luvata.com/Products/Hollow-Conductors}} (\bibinfo {year}
  {2013})\BibitemShut {NoStop}%
\bibitem [{\citenamefont {Barnes}\ \emph {et~al.}(2007)\citenamefont {Barnes},
  \citenamefont {Caspers}, \citenamefont {Ducimeti\`{e}re}, \citenamefont
  {Garrel},\ and\ \citenamefont {Kroyer}}]{PAC07-Barnes}%
  \BibitemOpen
  \bibfield  {author} {\bibinfo {author} {\bibfnamefont {M.~J.}\ \bibnamefont
  {Barnes}}, \bibinfo {author} {\bibfnamefont {F.}~\bibnamefont {Caspers}},
  \bibinfo {author} {\bibfnamefont {L.}~\bibnamefont {Ducimeti\`{e}re}},
  \bibinfo {author} {\bibfnamefont {N.}~\bibnamefont {Garrel}}, \ and\ \bibinfo
  {author} {\bibfnamefont {T.}~\bibnamefont {Kroyer}},\ }in\ \href@noop {}
  {\emph {\bibinfo {booktitle} {PAC'07}}}\ (\bibinfo {address} {Albuquerque,
  USA},\ \bibinfo {year} {2007})\ pp.\ \bibinfo {pages}
  {1574--1576}\BibitemShut {NoStop}%
\bibitem [{\citenamefont {He}\ \emph {et~al.}(2002)\citenamefont {He},
  \citenamefont {Hseuh},\ and\ \citenamefont {Todd}}]{He2002}%
  \BibitemOpen
  \bibfield  {author} {\bibinfo {author} {\bibfnamefont {P.}~\bibnamefont
  {He}}, \bibinfo {author} {\bibfnamefont {H.}~\bibnamefont {Hseuh}}, \ and\
  \bibinfo {author} {\bibfnamefont {R.}~\bibnamefont {Todd}},\ }\href {\doibase
  10.1016/S0040-6090(02)00661-2} {\bibfield  {journal} {\bibinfo  {journal}
  {Thin Solid Films}\ }\textbf {\bibinfo {volume} {420-421}},\ \bibinfo {pages}
  {38} (\bibinfo {year} {2002})}\BibitemShut {NoStop}%
\bibitem [{\citenamefont {Pont}\ \emph {et~al.}(2011)\citenamefont {Pont},
  \citenamefont {Nunez},\ and\ \citenamefont {Huttel}}]{IPAC11-Pont}%
  \BibitemOpen
  \bibfield  {author} {\bibinfo {author} {\bibfnamefont {M.}~\bibnamefont
  {Pont}}, \bibinfo {author} {\bibfnamefont {R.}~\bibnamefont {Nunez}}, \ and\
  \bibinfo {author} {\bibfnamefont {E.}~\bibnamefont {Huttel}},\ }in\
  \href@noop {} {\emph {\bibinfo {booktitle} {IPAC'11}}}\ (\bibinfo {address}
  {San Sebastian, Spain},\ \bibinfo {year} {2011})\ pp.\ \bibinfo {pages}
  {2421--2423}\BibitemShut {NoStop}%
\bibitem [{\citenamefont {Ghodke}\ \emph {et~al.}(2001)\citenamefont {Ghodke},
  \citenamefont {Angal-Kalinin},\ and\ \citenamefont {Singh}}]{APAC01-Ghodke}%
  \BibitemOpen
  \bibfield  {author} {\bibinfo {author} {\bibfnamefont {A.~D.}\ \bibnamefont
  {Ghodke}}, \bibinfo {author} {\bibfnamefont {D.}~\bibnamefont
  {Angal-Kalinin}}, \ and\ \bibinfo {author} {\bibfnamefont {G.}~\bibnamefont
  {Singh}},\ }in\ \href@noop {} {\emph {\bibinfo {booktitle} {APAC'01}}}\
  (\bibinfo {address} {Beijing, China},\ \bibinfo {year} {2001})\ pp.\ \bibinfo
  {pages} {363--365}\BibitemShut {NoStop}%
\bibitem [{\citenamefont {Barnes}\ \emph {et~al.}(2012)\citenamefont {Barnes},
  \citenamefont {Fowler}, \citenamefont {Atanasov}, \citenamefont {Kramer},\
  and\ \citenamefont {Stadlbauer}}]{IPAC12-Barnes}%
  \BibitemOpen
  \bibfield  {author} {\bibinfo {author} {\bibfnamefont {M.~J.}\ \bibnamefont
  {Barnes}}, \bibinfo {author} {\bibfnamefont {T.}~\bibnamefont {Fowler}},
  \bibinfo {author} {\bibfnamefont {M.~G.}\ \bibnamefont {Atanasov}}, \bibinfo
  {author} {\bibfnamefont {T.}~\bibnamefont {Kramer}}, \ and\ \bibinfo {author}
  {\bibfnamefont {T.}~\bibnamefont {Stadlbauer}},\ }in\ \href@noop {} {\emph
  {\bibinfo {booktitle} {IPAC'12}}}\ (\bibinfo {address} {New Orleans, USA},\
  \bibinfo {year} {2012})\ pp.\ \bibinfo {pages} {3686--3688}\BibitemShut
  {NoStop}%
\bibitem [{\citenamefont {{Infineon Technologies AG, Neubiberg,
  Germany}}(2018)}]{Infineon}%
  \BibitemOpen
  \bibfield  {author} {\bibinfo {author} {\bibnamefont {{Infineon Technologies
  AG, Neubiberg, Germany}}},\ }\href@noop {} {}\bibinfo {howpublished}
  {\url{https://www.infineon.com/cms/en/product/power/igbt/igbt-modules/igbt-modules-up-to-4500v-6500v/fd500r65ke3-k/}}
  (\bibinfo {year} {2018})\BibitemShut {NoStop}%
\bibitem [{\citenamefont {{Infineon Technologies AG, Neubiberg,
  Germany}}(2013)}]{InfineonSW3}%
  \BibitemOpen
  \bibfield  {author} {\bibinfo {author} {\bibnamefont {{Infineon Technologies
  AG, Neubiberg, Germany}}},\ }\href@noop {} {}\bibinfo {howpublished}
  {\url{https://www.infineon.com/cms/en/product/power/igbt/igbt-modules/igbt-modules-up-to-1200v/fd1400r12ip4d/}}
  (\bibinfo {year} {2013})\BibitemShut {NoStop}%
\bibitem [{\citenamefont {{The Okonite Company, 102 Hilltop Road Ramsey, New
  Jersey, USA}}(2013)}]{Okonite}%
  \BibitemOpen
  \bibfield  {author} {\bibinfo {author} {\bibnamefont {{The Okonite Company,
  102 Hilltop Road Ramsey, New Jersey, USA}}},\ }\href@noop {} {}\bibinfo
  {howpublished} {\url{www.okonite.com/}} (\bibinfo {year} {2013})\BibitemShut
  {NoStop}%
\bibitem [{\citenamefont {{Danfysik A/S, Gregersensvej 8, DK-2630 Taastrup,
  Denmark}}(2013)}]{Danfysik}%
  \BibitemOpen
  \bibfield  {author} {\bibinfo {author} {\bibnamefont {{Danfysik A/S,
  Gregersensvej 8, DK-2630 Taastrup, Denmark}}},\ }\href@noop {} {}\bibinfo
  {howpublished} {\url{www.danfysik.com/}} (\bibinfo {year} {2013})\BibitemShut
  {NoStop}%
\bibitem [{\citenamefont {{Tektronix, Beaverton, Oregon, United
  States}}(2014)}]{Tektronix:XYZs}%
  \BibitemOpen
  \bibfield  {author} {\bibinfo {author} {\bibnamefont {{Tektronix, Beaverton,
  Oregon, United States}}},\ }\href@noop {} {}\bibinfo {howpublished}
  {\url{https://www.tek.com/document/primer/xyzs-oscilloscopes-primer-1}}
  (\bibinfo {year} {2014})\BibitemShut {NoStop}%
\bibitem [{\citenamefont {{EPICS -- Experimental Physics and Industrial Control
  System}}()}]{bib:epics}%
  \BibitemOpen
  \bibfield  {author} {\bibinfo {author} {\bibnamefont {{EPICS -- Experimental
  Physics and Industrial Control System}}},\ }\href@noop {} {}\bibinfo
  {howpublished} {\url{epics-controls.org}},\ \bibinfo {note} {accessed:
  2019-05-19}\BibitemShut {NoStop}%
\bibitem [{\citenamefont {Masuda}\ \emph {et~al.}(2012)\citenamefont {Masuda},
  \citenamefont {Hatanaka}, \citenamefont {Jeong}, \citenamefont {Kawasaki},
  \citenamefont {Matsumiya}, \citenamefont {Matsuta}, \citenamefont {Mihara},\
  and\ \citenamefont {Watanabe}}]{PhysRevLett.108.134801}%
  \BibitemOpen
  \bibfield  {author} {\bibinfo {author} {\bibfnamefont {Y.}~\bibnamefont
  {Masuda}}, \bibinfo {author} {\bibfnamefont {K.}~\bibnamefont {Hatanaka}},
  \bibinfo {author} {\bibfnamefont {S.-C.}\ \bibnamefont {Jeong}}, \bibinfo
  {author} {\bibfnamefont {S.}~\bibnamefont {Kawasaki}}, \bibinfo {author}
  {\bibfnamefont {R.}~\bibnamefont {Matsumiya}}, \bibinfo {author}
  {\bibfnamefont {K.}~\bibnamefont {Matsuta}}, \bibinfo {author} {\bibfnamefont
  {M.}~\bibnamefont {Mihara}}, \ and\ \bibinfo {author} {\bibfnamefont
  {Y.}~\bibnamefont {Watanabe}},\ }\href {\doibase
  10.1103/PhysRevLett.108.134801} {\bibfield  {journal} {\bibinfo  {journal}
  {Phys. Rev. Lett.}\ }\textbf {\bibinfo {volume} {108}},\ \bibinfo {pages}
  {134801} (\bibinfo {year} {2012})}\BibitemShut {NoStop}%
\bibitem [{\citenamefont {Rawnsley}(2017)}]{bib:rawnsley}%
  \BibitemOpen
  \bibfield  {author} {\bibinfo {author} {\bibfnamefont {W.~R.}\ \bibnamefont
  {Rawnsley}},\ }\href@noop {} {\enquote {\bibinfo {title} {Ucn toroid
  processor development},}\ }\bibinfo {howpublished} {TRIUMF Internal Note}
  (\bibinfo {year} {2017})\BibitemShut {NoStop}%
\end{thebibliography}%

\end{document}